\documentclass[11pt,preprint]{aastex}

\def\sun{\hbox{$\odot$}}

\def\lesssim{\mathrel{\hbox{\rlap{\hbox{%
 \lower4pt\hbox{$\sim$}}}\hbox{$<$}}}}
\def\gtrsim{\mathrel{\hbox{\rlap{\hbox{%
 \lower4pt\hbox{$\sim$}}}\hbox{$>$}}}}

\def\arcsec{\hbox{$^{\prime\prime \, }$}}

\def\micron{\hbox{$\mu$m}}

\begin{document}

\title{Variability and Multiwavelength Detected AGN in the GOODS Fields} 

\author{Vicki L. Sarajedini}
\affil{University of Florida, Department of Astronomy, Gainesville, FL 32611; 
vicki@astro.ufl.edu}
\author{David C. Koo\altaffilmark{1}, 
Alison J. Klesman\altaffilmark{2}, 
Elise S. Laird\altaffilmark{3},
Pablo G. Perez Gonzalez\altaffilmark{4}\altaffilmark{5}
Mark Mozena\altaffilmark{1}}
\altaffiltext{1}{UCO/Lick Observatory, University of California, Santa Cruz, CA 95064}
\altaffiltext{3}{Astrophysics Group, Imperial College London, London, SW7 2AZ, UK}
\altaffiltext{2}{University of Florida, Gainesville, FL 32611}
\altaffiltext{4}{Departamento de Astrof\'{\i}sica, Facultad de CC. F\'{\i}sicas,
Universidad Complutense de Madrid, E-28040 Madrid, Spain}
\altaffiltext{5}{Associate Astronomer at Steward Observatory, The University of Arizona} 

\begin{abstract}
  
We identify 85 variable galaxies in the GOODS North and South fields using 
5 epochs of HST ACS V-band (F606W) images spanning 6 months.  
The variables are identified through significant flux
changes in the galaxy's nucleus and represent $\sim$2\% of the survey galaxies.
With the aim of studying the active galaxy population in the
GOODS fields, we compare the variability-selected sample with X-ray and mid-IR
AGN candidates.  Forty-nine percent of the variables
are associated with X-ray sources identified in the 2Ms Chandra surveys.  Twenty-four
percent of X-ray sources likely to be AGN are optical variables and this percentage increases with
decreasing hardness ratio of the X-ray emission.  Stacking of the non-X-ray detected variables
reveals marginally significant soft X-ray emission.  Forty-eight percent of mid-IR power-law sources
are optical variables, all but one of which are also X-ray detected.  Thus, about half of
the optical variables are associated with either X-ray or mid-IR power-law emission. 
The slope of the power-law fit
through the Spitzer IRAC bands indicates that two-thirds of the variables have BLAGN-like SEDs.  
Among those galaxies spectroscopically identified
as AGN, we observe variability in 74\% of broad-line AGNs and 15\% of NLAGNs.
The variables are found in galaxies extending to z$\sim$3.6.
We compare the variable galaxy colors and magnitudes to the X-ray and mid-IR sample and find 
that the non-X-ray detected variable hosts extend to bluer colors and fainter intrinsic magnitudes.
The variable AGN candidates have Eddington ratios similar to those of X-ray selected AGN.

\end{abstract}

\keywords{galaxies:active--surveys}

\section{INTRODUCTION}

Active Galactic Nuclei (AGNs) are galaxies accreting significant amounts
of material onto their central supermassive black holes (SMBHs).  These galaxies were
once thought to be oddities among the total galaxy population, but are now seen as
important mile markers on the broad road of galaxy evolution.  Once considered rare,
SMBHs are now believed to exist in the centers of all galaxies containing a significant
bulge component (Kormendy \& Richstone 1995).  
In addition, the observed relation between the mass of
the SMBH and the bulge stellar mass (Ferrarese \& Merritt 2000; Gebhardt et al. 2000) 
suggests a coupled assembly history between the two.
To understand the role played by AGN in the evolution of galaxies, it is thus necessary to 
identify complete samples of AGN in galaxy surveys.

AGN have been identified using many techniques such as UV/optical color selection (Markarian 1967;
Schmidt \& Green 1983; Richards et al. 2002),
spectroscopic signatures of broad emission lines or narrow emission lines with distinct
line flux ratios (Baldwin et al. 1981; Veilleux \& Osterbrock 1987), 
X-ray emission (Fabbiano et al. 1992; Alexander et al. 2003), 
infrared color selection (Lacy et al. 2004; Stern et al. 2005), 
radio emission (e.g. Smith \& Wright 1980) and 
variability (e.g. Hook et al. 1994; MacLeod et al. 2010)
over a range of timescales.  Many of these techniques, however, produce samples biased against
galaxies where the AGN light is much less than that of the host galaxy.  Additionally, most
surveys in the optical/UV are biassed against obscured, dusty AGN/host galaxies.   In order
to identify largely complete samples of AGN, it is clear that a variety of techniques and
selection methods should be employed.  While it may be found that AGN selected using different 
techniques represent separate populations experiencing different phases in AGN/galaxy evolution 
(e.g. Hickox et al. 2009), a less biased sample of AGN will help to make this picture clearer and
allow us to better interpret the role AGN play within their host galaxies.

In this paper, we aim to identify AGN in the Great Observatories Origins Deep Survey (GOODS)
North and South fields beginning with an optical variability selected sample.  AGN are known to vary
on timescales of months to years in the optical with 90--100\%
of AGN identified via other means observed to vary over the course of several years 
(e.g. Koo et al. 1986; Schmidt et al. 2010).
The mechanism behind the variability is still uncertain, though the leading explanation involves
disk instabilities (Pereyra et al. 2006) or changes in the amount of infalling material 
(Hopkins et al. 2006).  Regardless of the physics behind variability, AGN are observed to display
brightness variations of several percent over many years.  There is additional evidence of an increase
in variability amplitude for intrinsically fainter AGN (Bershady et al. 1998; Vanden Berk et al. 2004) 
making variability a particularly effective means for identifying low-luminosity AGN. 
We build on the results of previous successful surveys to identify AGN candidates in the nuclei
of galaxies found in deep HST surveys (e.g. Sarajedini et al. 2003).  The high resolution
HST images allow us to obtain accurate photometry within small apertures (r$\lesssim$0.2$\arcsec$),
thus allowing us to probe lower AGN/host galaxy luminosity ratios than can be done using ground-based
images.  We expect $\sim$60\% of AGN in the GOODS fields to be identified
as variable with the time sampling of 5 epochs separated
by 45-day intervals.  This estimate is based on the results of Webb \& Malkan (2000), who found
that 60\% of Seyfert/QSOs in their study revealed significant variability on month-to-month
timescales.

We also explore X-ray and mid-IR selected AGN identified in the GOODS fields and compare
with the variability survey.  These wavelengths are powerful probes of AGN and are especially
important for uncovering obscured sources.  
The multiwavelength data provide 
important confirmations of the AGN nature of
the variables as well as the quantification of the use of optical variability
in identifying sources of varying levels of obscuration.   We also compare the
variables to spectroscopically selected AGN in the GOODS fields.

We describe the AGN sample drawn from the variability analysis in \S 2.  In \S 3 we compare
this sample to those identified via X-rays, mid-IR power-law SEDs, and optical spectroscopy.
We discuss the redshift distribution, absolute magnitudes, and colors of the variables, X-ray
and mid-IR AGN in \S 4.  In \S 5 we discuss how variability selected AGN fit with the current
picture of AGN/galaxy evolution in part by deriving Eddington ratios for the variables
in our survey.  Throughout the paper we assume a flat, cosmological constant-dominated
cosmology with parameter values $\Omega$$_{\Lambda}$ = 0.7,  $\Omega$$_M$ = 0.3, and H$_o$ = 70
km/s/Mpc.  The data presented in this paper are in the AB photometric system.

\section{DATA ANALYSIS AND SELECTION OF VARIABLES}

The GOODS fields consist of two regions of the sky, each $\sim$130 arcmin$^2$ in size and located
at RA = 3h32m, Dec = -27d48m and RA = 12h37m, Dec = +62d15m.
These fields were imaged with HST's Advance Camera for Surveys Wide Field Camera (ACS WFC) in 
four passbands (F435W, F606W, F775W, and F850LP).  All
but the B-band images (F435W) were obtained in epochs separated by 45 days.
We chose to perform our analysis using the F606W (hereafter, V-band) images to
maximize sensitivity to varying AGN within host galaxies, given the fact that
several studies find increased variability
amplitude with decreasing wavelength (di Clemente et al. 1996;
Sarajedini et al. 2003).  The South field images
were obtained between August 1 2002 and February 5 2003 with 5
epochs separated by $\sim$45 days for a total time baseline of 6 months.
The North field images were obtained between November 21 2002 and May 23 2003
with the same time sampling as in the south.  Two exposures per pointing
per epoch were combined to produce an image with a total exposure time
of $\sim$1000s. 

The GOODS v2.0 source catalog
${(archive.stsci.edu/pub/hlsp/goods/catalog\_r2)}$
provided initial coordinate positions for identifying objects in the individual epoch ACS images.
We found 11519 sources in the combined fields that were visible in all 5 epochs.
To ensure that the centers of all sources were accurately determined
and matched across the epochs, objects were recentered using the IRAF task ${\it center}$
and sources whose centers had large pixel shifts
(i.e. more than 3$\sigma$ of the average shift for sources in a given ACS image)
were visually inspected.  Sources that fell too close to the CCD edge or
with a cosmic ray close to the object center in one or more epochs were rejected. 
After this process, 10161 sources remained.

We used IRAF ${\it phot}$ to determine aperture photometry for each source using
3 aperture sizes, r$=$2.5 pixels, 3.5 pixels, and 5 pixels where the pixel scale
is 0.05 arcsec/pix.  The smallest aperture size is
$\sim$twice the FWHM of a typical unresolved source and was initially
selected as the primary aperture to identify AGN in galaxies,
which should originate from an unresolved nuclear region.  However, we
carry out the variability analysis using the 5 pixel aperture data
to ensure that changes in the PSF, which would be more significant for smaller 
apertures, do not result in an overestimate of variables.
Magnitude differences were determined between epochs 2 through 5 and the first 
epoch.  Small zeropoint offsets of 
0.01 to 0.03 magnitude were determined among the epochs and applied to the photometry.
For every source, we then determined the mean magnitude and standard deviation 
around the mean.  We also calculate the concentration index, CI, as the magnitude difference between
the r$=$2.5 pixel photometry and the r$=$5 pixel photometry.  Point-like, unresolved
sources should have small CI values while extended sources will occupy a range of higher 
values.

Figure 1 shows the mean nuclear V-band magnitude versus 
standard deviation for r$=$2.5 pixel apertures (Fig 1a),
r$=$5 pixel apertures (Fig 1b), and the concentration index (Fig 1c).
The expected increase in
standard deviation values at fainter magnitudes is observed due to increasing
photometric noise.  The spread in standard deviation values is generally the same
using either the small or large aperture photometry and would suggest that variations in the
PSF are not producing significantly increased standard deviation scatter.
Figure 1c shows a clear separation between the resolved and unresolved sources.
Based on this figure, we define objects having CI values of $\leq$0.2 as unresolved 
or point-like and those with higher values as resolved or extended.

Applying this division, we find that unresolved
sources generally have higher standard deviation values than do the extended 
sources across the range of magnitudes (Figure 2).  The scatter in the standard 
deviation values do not decrease significantly when using the r$=$5 pixel 
photometry.  We analyzed the unresolved sources using r$=$10 pixel aperture 
photometry as well and found a similar level of scatter.
This higher level of scatter among the standard deviation values of
the unresolved population is likely reflecting the
difficulty in achieving photometric repeatability for point-like sources and may
be enhanced by variations in the PSF
which are not as significant for extended sources.  Since simply using larger 
apertures does not reduce the scatter, we account for this difference 
by continuing the variability analysis separately for the unresolved
and extended sources throughout the remainder of this paper.  In this way, we determine
the variability threshold for unresolved sources independently and can allow for
this larger scatter.  We use small
apertures (r$=$2.5 pixel) for the extended sources to isolate the varying nuclei of galaxies and
increase sensitivity.  For the unresolved sources, we use the
r$=$5 pixel apertures to avoid any possible sensitivity to PSF changes.
In addition, since these are generally point-like in morphology, 
there is no need to use the smaller
aperture to block light from an extended galaxy component.

We apply both a bright and faint magnitude cut to the variability survey.
A bright limit of V$_{2.5}=$20.0 was applied to remove several marginally saturated
unresolved objects at the bright end of the distribution.  These appear to be
mainly foreground stars for which accurate photometry can not be determined.
We identify the faint magnitude limit for our variability survey from the source
number counts as a function of magnitude.  The number of objects significantly drops 
beyond V$_{2.5}=$26.5.  Therefore, we limit the survey analysis to 
V$_{2.5}=$26.0 to avoid significant photometric incompleteness.

To identify the variability threshold, the standard deviation values 
are separated into magnitude
bins, allowing the bin limits to vary in size with larger bins at the bright
end and smaller bins at the faint end to maintain similar numbers of objects
ranging from 15 to 40 sources per bin.  The greatest outliers from the mean magnitude of
each bin were excluded and the resulting histograms were fit with Gaussians 
to determine the distribution center (mean standard deviation) and the sigma 
(i.e. spread) in standard deviation values at a given magnitude.  At fainter magnitudes
the standard deviation increases as well as the spread in those values.  We
fit a 2nd-order polynomial to both the mean standard deviation values in
each bin and 1$\sigma$ of the Gaussian distribution in each bin.  
This is done in order to characterize the photometric error as a smoothly
changing function of the source magnitude.  Both the mean of the standard deviation
and the spread in values around that mean are determined independently.
Figures 3a and b show the 
variability plot for the GOODS South field with the mean standard
deviation (solid thick line) and the mean plus 2-sigma and plus 3-sigma (dashed lines) for
the point sources and extended sources separately.
Figures 3c and 3d are the same for the North field.
Determining the expected error at each magnitude in this empirical way provides the most
realistic estimation of the photometric error on the standard deviation since we
carry through the effects of object and sky Poisson noise, readout noise, and any
limitations imposed by minor errors in the sky zero points used.

Sources at or above the mean standard deviation plus 3$\sigma$ were
flagged for visual inspection.  In the south field, 126 extended
sources and 27 point sources were flagged and 94 extended and 10 point sources
in the north field.  Visual inspection of these outliers revealed that $\sim$50 
in each field
contained evidence of cosmic rays or fell in a location too close to the edge of 
a CCD and thus suffered from high background noise in one or more epochs. 
These sources were removed from the survey.
About 15 objects per field were determined to have centering problems and these 
were manually recentered.
With all outliers checked for accuracy, we recalculated the 
variability threshold for each field. 

After comparing the variables against the known supernovae catalogs of
Strolger et al. (2004), we identified one SN in the north and one in the
south field from our extended source list.  We also compare to several published
redshift surveys (see \S 4) 
and identify a number of spectroscopically confirmed stars
(11 in each of the fields).  Since the aim of our paper to
study the AGN population in these fields, we remove the known SNe and stars from the survey. 

The final sample of galaxies surveyed for AGN variability is shown in Figure 4.
We normalize the
y-axis by dividing the standard deviation of each galaxy (with the mean standard 
deviation at that magnitude subtracted) by the 1$\sigma$ Gaussian fit  
determined for the galaxies at each magnitude, the majority of which are assumed to
be non-varying.  We call this the variability significance parameter.  The
variability significance will have a value of
3 for objects that are varying by exactly 3 times the 1$\sigma$ Gaussian
fit (i.e. spread) in standard deviation values at a given magnitude.
Since we subtract off the mean standard deviation, some sources will have negative
variability significance values.

To determine the threshold of variability and estimate the number of false positives,
we employ an empirical approach. 
We examine the cumulative distribution of variability signficance values
for resolved and unresolved galaxies in GOODS (Figure 5).  
For a non-varying population, a smooth decline in the distribution should be observed.
A population of significant outliers should appear as a break in that distribution with
a shallower decline at higher variability significance values.  A fit to the distribution
of the non-varying sources (long dashed line) helps to identify the break
at a significance value of $\sim$3.1 for the extended sources and at $\sim$3 for the 
unresolved sources (short dashed line).  We use this value for the threshold of
variability and indicate it with a dashed line in Figure 4.
To estimate the number of outliers,
we extend the fit to the smooth distribution of non-varying galaxies at low significance
values to the y-intercept. Integrating below this line and beyond the
variability threshold (shaded region in Figure 5) yields an estimate of the
expected number of outliers from the non-varying population.  With a total of
3775 extended sources and 399 unresolved sources, the shaded regions represent $\sim$4 
false positives among the extended and $\sim$3 false positives among 
the unresolved sources for a total of 7 expected false positive variables.

Based on our determined variability thresholds, 
we identify 42 variable galaxies out of a possible 
2055 galaxies in the GOODS South field and 43 variables out of 2119 galaxies in the GOODS
North field.  This represents 2\% of all galaxies within the flux limits of the survey
that display significantly varying nuclei over the 6 month time interval.
We estimate that $\sim$7 of the 85 total variables, or $\sim$8\%, are false positives.
Table 1 lists the variables identified in the GOODS survey.

\subsection{Comparison with Other Variability Surveys in GOODS}

Previous studies to identify variables in the GOODS fields or its subregions have been
published (Sarajedini et al. 2003; Cohen et al. 2006; Klesman \& Sarajedini 2007;
Trevese et al. 2008, Villforth et al. 2010).  Here we compare our results with Trevese et al. (2008)
and Villforth et al. (2010), the two studies
with the largest field overlap and data similarities.  We discuss
Klesman \& Sarajedini (2007) in \S 3 since that study consisted of
X-ray and mid-IR pre-selected AGN candidates which we consider in that section of the paper.

The survey with the greatest similarity
in terms of photometric data and field coverage is that recently published by
Villforth et al. (2010).  They identify a variability selected sample of
galaxies using the z-band ACS 5 epoch
GOODS images.  The sample of 139 likely AGN candidates are chosen based upon
significant variability using the $C$ statistic, a technique that compares
the observed standard deviation to the expected standard deviation for each source.
We find that 86 of their 139 candidates are also analyzed for variability in our
survey.  Those not included in our survey were removed due to the presence of a 
cosmic ray in or near the nucleus in one or
more epochs, the galaxy location being too close to the CCD edge in one or more
epochs, or not falling within the magnitude parameters of our survey as defined above.  
For those sources analyzed in both surveys, we find 23 variables in common or 27\%.
Villforth et al. divide their sample into a ``clean" sample and ``normal" sample,
where the clean sample consists of sources with $\geq$ 99.99\% variability
significance and the "normal" sample contains additional sources down to
99.9\% significance.  Twenty-two of the variables found in common are in the Villforth
et al. (2010) ``clean" sample and only 1 is found in the remaining ``normal"
sample.  Considering only the clean sample, 57 of their variables are in our 
survey and 22 were found to be variable.  Thus the percentage of variables
in common increases from 27\% to 39\% if we consider only the most significant variables
in their survey.

The sample differences may result from a number of
factors.  First, the difference in the wavelength of the filter used to identify variables may
play a role.  We found a larger number of variables and a greater overlap with X-ray
surveys in the HDF when analyzing variability in the V-band (Sarajedini et al. 2003)
as compared to the I-band (Sarajedini et al. 2000).  This is consistent with findings
that variability amplitude increases with decreasing wavelength 
(e.g. Vanden Berk et al. 2004).  Secondly, differences in the photometric aperture choice
used to identify
variability may also be important.  We use smaller apertures for the extended sources 
(r$=$0.125$\arcsec$ versus their r$=$0.36$\arcsec$)
which should be more sensitive to nuclear flux changes while still remaining largely 
unaffected by PSF variations (see Villforth et al. 2010, Figure 1).  
The aperture we use for unresolved sources is actually closer to that adopted by
Villforth et al. (2010).  We chose a larger aperture for these objects 
(r$=$0.25$\arcsec$ and additional tests at r$=$0.5$\arcsec$) to limit effects 
from PSF variations.
Additional comparisions of these samples are the focus of a future paper.

Another variability study in this field is that of
Trevese et al. (2008) in the Chandra Deep Field
South which encompases the GOODS-S.  The variables were selected from
a V-band ground-based survey using the ESO/MPI 2.2m telescope.  
The observations cover a time baseline of $\sim$2 years and are sampled
every few months.  They identify 132 variables in this larger field, of which
22 are in the ACS field-of-view.  Fourteen of these are included within the parameters
of our variability survey.  Of these 14, we find 8 as significant variables
in our survey or 57\%.  If we restrict the sample to those which Trevese et al. identify
as BLAGNs, we find 80\% as variable (8 out of 10).  

We identify a higher density of variables in our common field
(42 as compared to 14) which is likely due to the differences in image 
resolution.  The high resolution HST images allow us to use small aperture
photometry to isolate the varying galaxy nucleus and should be more sensitive to varying
AGN.  In addition, the image quality and repeatability of the HST data 
allows for the detection of 
variability to a lower threshold in terms of standard deviation.   The variables identified
by Trevese et al. (2008) that are not found in our survey may result from longer
and more complete time sampling of the AGN light curve.  Our survey covers a 6 month
time interval and is expected to be $\sim$60\% complete in detecting variable AGN while
longer time sampling over 1 to 2 years should identify virtually all varying AGN (e.g.
Koo et al. 1986; Schmidt et al. 2010).

\section{COMPARISON WITH MULTI-WAVELENGTH SURVEYS}

The GOODS North and South fields have been the target of deep X-ray 
observations with Chandra and mid-IR imaging with Spitzer IRAC and MIPS.  
We compare AGN candidates selected via X-ray detection and mid-IR power-law
SED fitting to our variability selected AGN candidates to investigate their
multiwavelength properties and various biases among the different selection
techiniques.  We also examine evidence for AGN based on emission line properties 
from published redshift surveys in these fields.  

\subsection{X-ray Detected AGN}

All optical sources surveyed for variability were matched
against the list of X-ray point source detections from the 2Ms Chandra Deep Field
surveys in the north field (Alexander et al. 2003) and the south (Luo et al. 2008).  
We refer to the ``variability survey" as those optical sources from the GOODS v2.0 
catalog that were visible in all
5 epochs with good quality images (i.e. containing no cosmic rays or CCD edge
effects) and having nuclear magnitudes brighter than V$_{2.5}$$=$26.

In the south field, Luo et al. (2008) found 218 optical matches to the 462
X-ray sources in the main catalog with a matching radius of 0.5$\arcsec$.
We find that 115 of these X-ray/optically matched sources fall within our variability survey. 
A supplementary catalog, produced with greater sensitivity in the outer regions
of the CDF-S due to the inclusion of $\sim$ 250 ks from the E-CDF-S, consists of
86 sources, of which 3 match to sources in the variability survey.  
A third catalog of bright optical sources with lower X-ray significance detections
consists of 30 sources, of which 15 have counterparts in the variability survey.
Although most objects in the optically bright catalog are not likely to be AGN based on their
low X-ray-to-optical flux ratios, we identify these matches since variability
would cast light on the nature of the X-ray emission. 

In the north field, Alexander et al. (2003) produced a main source catalog 
containing 503
sources in the CDF-N, and a bright source catalog containing 75 sources
with lower significance X-ray detections matched to optically bright galaxies.
Barger et al. (2003) identified optical matches and optical source positions for
the main source catalog objects based on ground-based Subaru Suprime-Cam observations
(Capak et al. 2003).  We match the GOODS-N v2.0 positions to these optical
positions rather than the X-ray positions for greater matching accuracy.   
There are 139 objects in the variability survey with matches to X-ray sources 
in the main catalog within 0.5$\arcsec$ of the optical counterpart position given in Barger
et al. (2003).  Two objects with offsets between 0.5 - 1.0$\arcsec$ were included
as matches since their X-ray positions matched the GOODS-N optical positions within 
$\sim$0.6$\arcsec$.  Among the bright source catalog, 49 objects match within 
0.5$\arcsec$ of an X-ray source.

We find 21 of the 42 variables in the south field 
and 21 of the 43 variables in the north field are X-ray detected,
yielding a total of 42/85 or 49\% of the variables associated 
with X-ray sources (blue triangles in Fig 4).  
This fraction is similar to that found in the HDF-N (Sarajedini et al. 2003)
and higher than recent ground-based optical variability surveys
(Travese et al. 2008; Morokuma  et al. 2008) where 60--70\% of variables
were not detected in X-rays.  
In Figure 6,
we plot the soft band flux (0.5 to 2 keV) against the R-band AB magnitude 
for all X-ray detected sources in the variability survey.
{\footnote{R-band magnitudes for the GOODS-S field are from Luo et al. (2008).  
GOODS-N magnitudes
are from Barger et al. (2003) for the main source catalog objects
and Alexander et al. (2003) for the bright source catalog.  The latter were 
converted from Vega magnitudes to AB by adding 0.2.}}
The solid lines indicate the location of most AGN with log($F_{x}/F_{opt}$) between -1
and 1 and the dashed line indicates log($F_{x}/F_{opt}$) = -2.  Most non-AGN X-ray sources
fall below the dashed line.  It can be seen that the majority of bright source
catalog objects (open circles) are not likely to be AGN based on their low $F_{x}/F_{opt}$
values.  The green points indicate the variable sources.  We
find that the majority of variables are likely to be AGN based on their location
in this diagram but a significant number (19\%) have low $F_{x}/F_{opt}$ values.
None of the objects in the X-ray bright source catalog for the GOODS-S field were
found to be variable and only one from the GOODS-N field (ID 2426) is variable.  
This object lies close to the limit of likely AGN $F_{x}/F_{opt}$ values.  

Figure 6 also shows the R-band magnitude for variables without an X-ray
counterpart (green asterisks).  These non-X-ray detected
variables cover a range of optical magnitudes but tend to be fainter than
the X-ray detected variables sources 
(average R$_{AB}=$22.1 for X-ray detected sources and 23.3 for
non-X-ray detected sources).  Thus, in general they are expected to be more difficult to detect
at the flux limits of the Chandra survey.  The range of optical magnitudes
for the non-X-ray detected sources would indicate that if the optical variability
is due to the presence of an AGN, they are likely to have low 
$F_{x}/F_{opt}$ values, similar to what is found 
for $\sim$20\% of the X-ray detected sources.  Additionally, since the optical
magnitudes represent the total galaxy+AGN light, all of the points on the diagram
will have brighter R-band magnitudes than for the AGN component alone, the extent 
of which depends on the ratio of AGN to host galaxy
luminosity.  Therefore, some of the X-ray non-detections, may be extremely
faint optical AGN that lie below the flux limits of the X-ray survey.

We explore the variable nature of the X-ray
detected sources and find that 21 of the 118 X-ray sources in GOODS-S 
(combining the main and extended CDF-S catalogs) are significant variables.  
In GOODS-N, 21 of the 140 X-ray sources are significant variables.  
Thus, 16.3\% of the X-ray sources with optical counterparts in the
variability survey are identified as significant variables.
We do not include the non-variable bright source catalog objects in
this statistic since the aim of our paper is to investigate the varying and 
non-varying AGN population.  With the same reasoning, we may also consider
X-ray sources below the dashed line in Figure 6 as less likely AGN candidates.  
Only 2 variables lie below this line in the north field (one very close to the limit)
and none in the south field.  If we exclude X-ray sources below log($F_{x}/F_{opt}$) = -2,
we find 21 of 75 X-ray sources in the south and 19 of 92 in the north are variable,
increasing the percentage of X-ray sources with optically varying counterparts
to 24\%.  This percentage is similar to that found in other variability studies.
Klesman \& Sarajedini (2007) found that 
26\% of the X-ray source population in the GOODS-S field were optical variables.  
Since this earlier analysis examined X-ray sources detected in the previously 
published 1Ms Chandra survey,
many of the fainter X-ray sources and those likely to have lower $F_{x}/F_{opt}$
values were already omitted from their sample (see discussion in Klesman \&
Sarajedini 2007).  Of the 21 objects we identify as optically variable X-ray sources,
18 were also identified in Klesman \& Sarajedini (2007). Of the three that were
not identified in that paper, one was not included in their analysis since it was 
only detected in the 2Ms survey.  The other two fell just below their variability significance
threshold. 
  
We expect that some X-ray detected sources do not reveal significant
variability in their optical counterparts due to obscuration.
We quantify obscuration using the hardness ratio,
HR, defined as the hard band source counts (HB; 2 - 7 keV) divided by the
soft band counts (SB; 0.5 - 2 keV).  Higher values of HR are found for more
obscured sources.  Figure 7 shows the hardness ratio vs. the variability
significance for all of the X-ray detected sources in our survey with enough
counts in either the soft or hard band.  Open points are upper limits without
enough counts in the hard band to compute a ratio.  
The solid line shows the threshold above which we consider the source to be 
significantly varying.  As in Klesman \& Sarajedini (2007), we find that
the majority of variables are "soft" sources, having HR values less than 0.5.
These are also the most significant variables in our sample.  We find that
36\% of the softest sources are variable, a dramatic increase from the 16\%
identified as variable overall.  The number of optical variables falls
significantly among harder, more obscured, X-ray sources.  For sources 
with HR values between 0.5 and 2, we find 8\% are optically variable.  
For the hardest sources with HR values greater than 2, only 1 variable is
found accounting for 3\% of the X-ray detected sources in our survey.
We observe that the variability significance declines
with increasing hardness ratio consistent with findings in Klesman \&
Sarajedini (2007).  

We continue to examine the non-X-ray detected variables by performing an
X-ray stacking analysis for these sources.  The stacking was done for the
full, soft and hard bands (Table 2).  The last column lists the S/N ratio
for the stacked samples.  Calculating the source significance instead, the
significance for the full, soft and hard bands for the GOODS-N is
1.6$\sigma$, 3.4$\sigma$, and 0$\sigma$, indicating a slightly significant
result only for the soft band.  For the GOODS-S field, source significance
is 2.5$\sigma$, 2.3$\sigma$ and 1.5$\sigma$.  The overall results of the analysis are
consistent with the trends observed in Figures 6 and 7, that optical variability 
is more likely to be detected among
X-ray soft sources and that many non-X-ray detected variables may emit
very weak X-rays and have F$_x$/F$_{opt}$ values lower than typical AGN.

The lack of X-ray emission from about half of the sources is not well understood, 
but is also observed in several previous variability surveys 
(Trevese et al. 2008, Morokuma et al. 2008).  Brandt \& Hasinger (2005) 
discuss some possible explanations (e.g. the lack of an accretion-disk X-ray corona) which may 
play a role in addition to the possibility that weak X-ray 
emission is present but not detectable at the current X-ray survey depths.
 
\subsection{IRAC Power-law Detected AGN}
Infrared selection of AGN is a powerful technique.  Several strategies
have been employed using mid-IR observations to identify AGN where some
portion of the light has been reprocessed by obsurring dust in the vacinity of 
the accreting blackhole (e.g. Lacy et al. 2004).  Donley et al. (2008)
have found that galaxies whose SEDs display power-law behavior in the
Spitzer IRAC 3.6 - 8$\micron$ bands represent the purest selection of
AGN using mid-IR observations, with many fewer contaminants than IRAC
color-color selection or IR-excess AGN selection techniques.  We have
thus compared our variability selected AGN candidates to the mid-IR
power-law galaxies identified in the CDF-N (Donley et al. 2007) and
the CDF-S (Alonso-Herrero et al. 2006) which overlap well with the
GOODS-N and GOODS-S optical observations. 

In the CDF-N, Donley et al. (2007) identified 62 power-law galaxies
from the Spitzer IRAC data.  Of these, 11 match to optical sources
in our variability survey to within 0.7$\arcsec$.  
The small overlap in samples is mainly due to the
difference in field size and the fact that some IR sources 
did not fall within the variability survey magnitude limits.
Of these 11, 4 are significant optical variables.  In the 
CDF-S, Alonso-Herrero et al. (2006) found 92 power-law galaxies.
Of these, 14 match to optical sources in our variability survey and 8 are significant 
optical variables.  In total we find
that 48\% of the mid-IR power-law galaxies in our survey reveal significant
optical variability.  This percentage is very similar to that found
by Klesman \& Sarajedini (2007) for the south field. 
Mid-IR power-law sources represent 14\% of the
85 variables identified in our survey.  

We can further study the nature of the optical variables in our survey
by investigating the slope of the power-law fit in the Spitzer IRAC 3.6 - 
8$\micron$ bands.
Power-law galaxies are selected as those sources with IRAC SEDs well
fit with f$_{\nu}$ $\sim$ $\nu$$^{\alpha}$ and having ${\alpha}$ $<$ -0.5.
Alonso-Herrero et al. (2006) classify power-law galaxies into catagories
based on the slope of the SED, broadly separating the BLAGN-like SEDs
from the NLAGN/ULIRG-like SEDs.  Steeper (i.e. more negative) values of
$\alpha$ more closely match templates of 
NLAGNs and ULIRGs.  Shallower SEDs resemble templates for BLAGNs.
These catagories separate at a spectral index of approximately $\alpha$$\simeq$-0.9.
From the published power-law slopes for our variables, we find that the majority of
mid-IR power-law galaxies which show significant variability
have spectral indices similar to BLAGN SEDs (8 out of
12) and those are also the most significantly varying sources. 
The other third of the variables have steeper SEDs through the IRAC channels
and thus would be catagorized as NLAGN or ULIRG-type SEDs. 

\subsection{Spectroscopically Identified AGN}
We compare the variability selected sample to spectroscopically
identified AGN in the literature.  For the north field, spectroscopic AGN
have been identified by Barger et al. (2008) and in the south by Santini et al. (2009).
Both papers are compilations of several redshift catalogs referenced within.
BLAGNs are classified as sources with clearly broad emission lines while NLAGNs 
are based on the presence of high ionization emission lines such as [NeV], CIV or CIII]1909.

From these spectroscopic catalogs, we find a total of
27 BLAGNs and 26 NLAGN galaxies that fall within our variability survey.  Of those, 20 of the BLAGNs
and 4 of the NLAGNs are identified as significant variables.  Thus, we find
variability in 74\% of broad-line AGNs and 15\% of narrow-line AGNs previously
observed in the GOODS north and south fields.  
Based on the total number of variables with optical spectroscpic data (see discussion below),
we find spectroscopic evidence of AGN in 40\% of the variables.

\section{GENERAL PROPERTIES OF AGN CANDIDATES AND HOST GALAXIES}

Both the north and south GOODS fields have been the target of extensive
spectroscopic follow-up. 
For the south field, we have the spectroscopic catalogs of Szokoly et al. (2004), Cimatti
et al. (2002), La Fevre et al. (2004), Vanzella et al. (2008), Mignoli et al. (2005) and
Popesso et al. (2009).  In the north, we have the compilation of spectroscopic
data in Barger et al. (2008) which includes the catalogs of Cowie et al. (2004), Reddy
et al. (2006) and Cohen et al. (2000, 2001).  
For sources without spectroscopic information,
photometric redshifts were obtained from the Rainbow Database (Perez-Gonzalez et al. 2008; 
Guillermo et al. 2010) for the south field and from Fernanez-Soto et al. (1999),
Capak et al. (2004), and Bundy et al. (2009) for the north.

Among the 4174 galaxies surveyed for AGN variability in the combined fields,
2371 have spectroscopic redshifts (57\%) and an additional 1497 have 
photometric redshifts yielding redshift information for 93\% of the galaxies.
Among the 85 variable galaxies, 62 have spectroscopic redshifts (73\%) and
the remaining 23 have photometric redshifts.

Figure 8a shows the redshift distribution for all galaxies in our survey compared
to the distribution of optical variables multiplied by 12 for comparison.  
The variables have a higher median
redshift than the total population with 35 lying at redshifts greater than 1.
The percentage of galaxies hosting optically varying nuclei
increases with increasing redshift, from $\lesssim$2\% at low redshifts to
almost 10\% at redshifts greater than 3 (Figure 8b). This is consistent with the
fact that the AGN luminosity function shows their numbers peak at redshifts between 
0.7$\lesssim$z$\lesssim$2, with more luminous AGN peaking at higher redshifts than
fainter AGN (e.g. Bongiorno et al. 2007).    

To further explore the nature of the AGN candidates, Figure 9 shows
redshift vs. rest frame absolute V magnitude
for all galaxies in our survey with the variable AGN candidates (green circles), X-ray detected
sources (blue triangles), and mid-IR sources (red squares) indicated.  We determine
rest-frame magnitudes using the v2.0 GOODS catalog b, v, i and z (F435W, F606W, F775W, and F850LP)
galaxy photometry (MAG-AUTO) and U-band photometry
from Wolf et al. (2004; GOODS-S) and Capak et al. (2004; GOODS-N).
These data, together with the redshift information, allows us to compute
k-corrected, rest-frame magnitudes and colors for each galaxy in our survey using 
$\emph{kcorrect}$ (Blanton et al. 2003).

Figure 9 reveals that AGN candidates generally have brighter
optical magnitudes than the non-AGN galaxies.  Their luminosity
may be partly enhanced by the AGN component, though AGN host galaxies are 
also typically found to be brighter and more massive than
galaxies not hosting AGN (e.g. Brusa et al. 2009).  
The X-ray detected AGN candidates
appear to lie in the brightest galaxies.  Mid-IR power-law galaxies
cover a range of magnitudes with most lying at the bright end and a
few occupying faint galaxies.  The mid-IR sources are also generally at
higher redshift than the X-ray sources, consistent with other studies of
X-ray and mid-IR AGN (Alonso-Herrero et al. 2007). 
Most optically bright variable AGN 
candidates are also X-ray sources.  The non-X-ray detected 
optical variables tend to have fainter absolute magnitudes at all redshifts.
Thus optical variability appears to identify some intrinisically faint
AGN/host galaxies and may do so more efficiently than X-ray or mid-IR
AGN identification.  

Figure 10 is the color-magnitude diagram for the galaxies in our survey
with the same symbols used to identify optical variables, X-ray, and mid-IR
sources.  We plot the rest-frame U -- V color versus the V-band
absolute magnitude for galaxies divided into two redshift bins, z=0.2$<$z$<$0.76 
and 0.76$<$z$<$1.6.  The second bin is more than twice the co-moving
volume of the first bin.  We see the bimodal color distribution observed in
recent surveys (e.g. Hogg et al. 2002) among our survey galaxies.
The dashed line indicates the limit of red sequence
galaxies as determined in Bell et al. (2004) at the center of each redshift bin.  

Several recent findings suggest that the integrated galaxy colors for the AGN candidates 
in our survey should largely represent the
colors of their host galaxies and in most cases are only marginally effected by the 
light from the nuclear AGN.  Recent results by Cardamone et al. (2010) for X-ray selected
AGN show little color contamination ($<$0.1 mag) from the nucleus in rest-frame
U -- V.  Likewise, Hickox et al. (2009) find a maximum of 0.4 -- 0.5 magnitude color corrections
are needed for the bluest AGN sources in (u -- r).  Exceptions are
the most luminous X-ray/optically variable AGN candidates, where we
expect the blue colors are indeed significantly affected by the AGN light.  These would
include several bright (M$_V$$<$-23) blue optical variables/X-ray sources in the 
lower left quadrant of Figure 10b. 

The AGN galaxy colors
cover a range in U--V and fill the region known as the green
valley, as observed previously for X-ray detected AGN 
(e.g. Nandra et al. 2007).  Figure 11 shows the color histograms for optical
variables, X-ray and mid-IR selected AGN candidates compared to the total galaxy
color population.  Here we include only galaxies at 0.2$<$z$<$1.6 and correct
for the slope of the dashed line in Figure 10.  We have also removed the 9 bright,
blue sources in the high redshift color magnitude diagram (Fig 10b)
that are likely to be affected by the AGN light and may not represent the true host galaxy colors.  
We find that the X-ray sources and optical variables occupy galaxies
at a range of colors, including many within the green valley.  While the
X-ray sources appear to peak in the green valley, the variables have a flatter
distribution that extends blueward of the X-ray host distribution and
into the red sequence galaxies.  There are few mid-IR AGN candidates at z$<$1.6, but
those in this redshift range show a relatively flat distribution.  This is
consistent with the results of Hickox et al. (2009) who find similar distributions
among a sample of 585 AGNs selected via X-ray and mid-IR colors.   
Recently, Cardamone et al. (2010) found that 75\% of X-ray hosts in the green valley
have colors consistent with young star-forming galaxies reddened by dust.  
The blue extension of the optically variable AGN hosts may imply 
that variability selected AGN are generally less dust reddened.  In any case, we
find a number of AGN hosts, several identified only through their variable nature, with blue colors 
indicative of current star formation.

\section{DISCUSSION}

We have shown that optical variability can be used to identify a significant population
of AGN candidates in deep HST fields.  The colors and magnitude ranges of the variable hosts
are similar to those of other AGN samples selected via X-rays or mid-IR
emission but extending to bluer colors and fainter magnitudes.  With about 50\% of 
the variables not identified as AGN through other means, we investigate
how the inclusion of this population fits with recent theories of
AGN/galaxy evolution.  

In the current picture of AGN and galaxy evolution, dark matter halos grow in the early
universe to masses of 10$^{12}$ to 10$^{13}$ M$_{\sun}$ when a dramatic event 
(e.g. major merger) triggers luminous quasar activity and 
rapid growth of the central black hole producing a dynamically hot stellar bulge component
in the host galaxy (Hopkins et al. 2008; 2009).  
Other events such as disk instabilities may also trigger accretion
onto the black hole and grow the stellar bulge.  After this period of growth, star 
formation in the galaxy 
must be quenched in order to produce the observed population of passively evolving galaxies
on the red sequence.  There has been speculation that the quenching results from the
event that fuels the quasar and/or is related to radiative feedback during the powerful 
AGN phase (Alexander et al. 2005; Brand et al. 2006).
The high virial temperature of the dark matter halo should stop accretion of cold gas
and thus inhibit further star formation (e.g. Faber et al. 2007).  The transition of
blue, star forming galaxy to a galaxy on the red sequence occurs relatively quickly
(1 -- 2 Gyr; Barger et al. 1996)
which is supported by the dichotomy of observed galaxy colors seen here and in other
surveys.  There is some evidence that further accretion onto the SMBH occurs during this
transition and would explain the observed "green'' colors of X-ray selected AGN
(Schawinski et al. 2007), though recent results suggest that many of those may 
actually be dust reddened star forming galaxies (Cardamone et al. 2010b).  
Galaxies on the red sequence may continue to experience
quenched star formation
due to mechanical feedback from the AGN which prevents the accretion of cool gas onto
the SMBH (Churazov et al. 2002). 

Optical variability is a ubiquitous observational characteristic
of accretion onto the SMBH and thus we might expect to identify AGN in the full range
of evolutionary stages described above.  However, since the variability 
likely originates from the UV/optically bright accretion disk, variability selection should be most sensitive 
to unobscured, less dust-reddened AGNs which is largely consistent 
with our findings.  The variables that are also X-ray and mid-IR sources
have luminosities, redshifts, and host galaxy colors consistent with 
the findings of Hickox et al. (2009).  Using the observed colors, clustering
properties, and SMBH accretion rates, they speculate that the X-ray
selected AGN observed at 0.25$<$z$<$0.8 live in large dark matter halo galaxies
that have recently experienced the buildup of their bulges and quenching of star formation.
These galaxies also exhibit a range of Eddington ratios (10$^{-3}$ to 1) which would
indicate declining accretion rates onto the SMBH.
The mid-IR AGNs in their study are associated with galaxies having less massive
dark matter halos and higher Eddington ratios than the X-ray selected sample.  Thus,
these galaxies have not reached the critical halo mass for the quenching of star formation
and therefore are more commonly found in blue host galaxies.  Though we do not
examine the clustering properties of the different samples here, the colors and magnitudes
of the X-ray selected and mid-IR selected variables in our survey would appear to be
consistent with this interpretation.  

To further investigate the variable AGN candidates found in our survey and attempt
to understand their placement in the above evolutionary scenario, we estimate the
black hole masses and accretion rates for the variability selected AGN candidates
at redshifts less than z$<$1.6.  Higher redshift data are excluded in this analysis
to avoid the high luminosity AGN which may contaminate galaxy colors
and bulge estimates for the host. 
To determine Eddington ratios however, we first need to estimate the luminosity of
the AGN.  We expect
the AGN component to be $\sim$10\% or less of the total galaxy flux in most cases.  This is
based on the observed variability in the nucleus compared with that expected from AGN structure function
flux amplitudes at time intervals similar to our data (e.g. Butler et al. 2010).
Nonetheless, we can make a crude estimate of the AGN flux by measuring the nuclear magnitude
within r=0.125$\arcsec$ apertures, twice the FWHM for the ACS images.  This aperture
contains the highest concentration of the AGN light and represents
an upper limit on the AGN luminosity since this measurement will contain essentially all of
the AGN plus some underlying flux from the host galaxy.
The nuclear V-band magnitudes are converted to rest-frame 
B-band absolute magnitudes.  We use the bolometric corrections of Hopkins et al. (2007)
to convert to bolometric luminosity.  Figure 12a shows the upper limit bolometric luminosity versus redshift
for the variables at z$<$1.6 where X-ray detected variables are shown as black points and
non-X-ray detected variables are green.
We find that the upper limit bolometric luminosities for non-X-ray detected
variables are generally fainter than that for the X-ray selected sample at all redshifts but with a large
dispersion.  

To estimate black hole masses, we compute the stellar bulge masses for our variable
galaxies.  We used GALFIT (Peng et al. 2002) to fit a 2-component model consisting
of an exponential disk and n$=$4 sersic profile bulge in both the V(F606W) and i-band(F775W) images.  
Fits with an n$=$2 sersic bulge
were also computed but showed no significant differences with the n$=$4 fits. 
With the magnitudes for each component determined,
we then calculated the total galaxy and bulge masses using the color and galaxy redshift.
Reliable bulge masses were calculated for 48 of the 62 variable galaxies at z$<$1.6 using this technique.
In some cases a significant bulge component was not identified in one of the bands, making
a bulge mass measurement impossible.  We use
the relation from Marconi \& Hunt (2003) to convert bulge mass to
black hole mass and plot this as a function of redshift in Figure 12b.
The variables have a range of black hole masses with an average of 7.2 log(M$_{BH}$[M$_{\sun}$]).
There is no apparent difference between the non-X-ray and X-ray detected variables at z$<$1.
At higher redshifts, the non-X-ray detected variables appear to have lower mass black holes in general,
though the number of sources at these redshifts with reliable bulge measurements is fewer.  We note 
that any significant contribution by the AGN component could produce an artificially high bulge mass 
and consequently overestimate the black hole mass.  While we expect the contribution of the AGN 
component to be negligible in the majority of our sources, the higher redshift and X-ray detected 
sources are more likely to be susceptible to this effect based on their overall colors and 
luminosities (as discussed in $\S$4).  This may account for the slightly higher BH masses 
for z$>$1 variables that are also X-ray detected in Figure 12b.

Finally, we compute the Eddington ratios using the equation
L$_{Edd}$ = 1.3 $\times$ 10$^{38}$ (M$_{BH}$/M$_{\sun}$) erg/s.  Figure 12c shows the Eddington
ratio as a function of redshift.  Several sources at z$>$1 have 
luminosities brighter than Eddington which is likely 
caused by the fact that our bolometric luminosity is an upper limit.  Thus, these values are less
reliable estimates of the true Eddington ratios of these sources.
Nonetheless, we find no discernible difference among the X-ray and non-X-ray
detected variable AGN across the range of observed redshifts.

These results suggest that the hosts of variability-detected AGN comprise a mix of galaxy-AGN types.
Some of the variables appear to be similar to other X-ray selected sources, having hosts with significant
stellar bulges and recently quenched star formation, 
followed by a decline in the accretion rate onto the black hole (Hickox et al. 2009).
In other cases the SMBHs may be continuously or stochastically fueled, possibly
at lower levels by stellar winds or bars in disk galaxies.  Such systems could have slightly
higher Eddington ratios and be in intermediate mass, later-type hosts (Hopkins et al. 2009) much
like the mid-IR/optical AGN discussed in Hickox et al. (2009).  This
scenario would be consistent with the observed lower-luminosity AGN within bluer host galaxies.

\section{CONCLUSIONS}

We have conducted a variability survey for AGN candidates in the GOODS fields and compared
this selection technique with results from other multiwavelength surveys of these fields.  
Using small aperture photometry, we identify 85 galaxies
with significantly varying nuclei in the HST ACS V-band images.  This represents 2\% of all
galaxies surveyed to a limiting magnitude of V$_{2.5}$=26 with an estimated 7 false positives.  
Forty-two of the 85 variables are associated with
X-ray sources detected in the Chandra 2Ms surveys of these fields and 14 of the
variables have power-law SEDs through the Spitzer IRAC mid-IR bands.   All but one of the mid-IR
sources is also an X-ray source, resulting in a total of 43 of 85 variables (51\%) with additional
multiwavelength 
evidence for the presence of an AGN.  Optical variability is more sensitive to soft X-ray
sources with fewer optical variables found to be associated with the harder X-ray detected 
sources in these fields.  
This is consistent with the results of X-ray stacking for the non-X-ray detected variables which
reveals a marginally significant detection in the soft X-ray band (0.5--2 keV) only.
We also find that the slope of the power-law fit through the mid-IR bands for optical variables
is consistent with BLAGN-like SEDs for 2/3rds of the variable AGN candidates with
the other third having steeper slopes like those for NLAGNs and ULIRGs.  Likewise, among
spectroscopically identified AGN, 74\% of BLAGNs and 15\% of NLAGNs show optical
variability.   

Our findings indicate that optical variability most efficiently
identifies less obscured, Type I - like AGN, though a surprisingly significant fraction (15 -- 30\%) 
appear to have some degree of obscuration revealed through the lack of an observed broad-line
region, slightly harder X-ray emission, or the slope of the mid-IR SED.  The overall number density
of variable AGN candidates is about half that of likely AGN X-ray sources with optical counterparts
in the GOODS fields and almost 3 times that of mid-IR power-law AGN with optical counterparts.
Thus, variability uncovers a significant population of AGN
which complements detection methods at other wavelengths.   

The variable AGN candidates span a broad range in redshift with the percentage of galaxies 
hosting variable nuclei increasing from $\lesssim$2\% at low redshift to $\sim$10\% at z$\sim$3. 
The total galaxy+AGN absolute magnitudes reveal that AGN candidates (variable, X-ray
and mid-IR detected candidates) are generally found in brighter galaxies at a range of colors spanning that of the
total galaxy population.  The color distribution of X-ray detected AGN hosts reveals a
peak in number density between the blue, star-forming and red sequence galaxies.  
The variables follow a similar trend but with an extension towards 
bluer colors and fainter absolute magnitudes.

Based on the derived bolometric luminosities, BH masses and Eddington ratios for the variable galaxies
in our survey, we find that non-X-ray detected variables at z$<$1.6 generally have lower luminosities 
than X-ray detected variables, but similar BH masses and Eddington ratios.  Thus, some of the variables
have characteristics like other X-ray selected samples, revealing significant 
bulges and the onset
of quenched star formation and decreased BH accretion rates, while others may be intermediate 
mass systems still forming stars and accreting at a continuous,
lower-level, possibly through stellar winds or bars.

The results of this survey suggest that variability is a promising technique to identify samples of AGN
extending to low-luminosities out to z$\sim$3.  These results are especially encouraging in
light of future planned multi-epoch observing programs like that for the LSST.  Observations with longer
and better sampled light curves will produce more robust and complete variability-selected AGN samples
in the low-luminosity, z$>$1 regime comparable to the well-quantified light curves and structure
functions for QSOs (e.g. MacLeod et al. 2010).
Together with multiwavelength surveys using a variety of detection techniques, the AGN population can be
more powerfully probed and understood in the context of galaxy evolution.

\acknowledgments
We thank Seth Cohen and Namish Hathi for providing a photometric redshift compilation for
the GOODS north field, Amy Barger for supplying the list of spectroscopic AGN identified in the
GOODS south field, Conor Mancone for assistance with k-correction of galaxy magnitudes, and
Philip Orange for work on the X-ray stacking analysis presented here.  We also thank the referee
for a careful reading of this paper and helpful comments and suggestions.

This paper is based on observations with the NASA/ESA Hubble Space Telescope, 
obtained at the Space Telescope Science Institute, which is operated by the 
Association of Universities for
Research in Astronomy, Inc., under NASA contract NAS5-26555
and observations with the Keck telescope, made possible by the W. M. Keck
Foundation and NASA.  Funding was provided by STScI grant AR-09948.01 and
NSF CAREER grant 0346691.

PGP-G acknowledges support from grants AYA2009-10368 and CSD2006-00070, and
the Ram\'on y Cajal Program, all financed by the Spanish Government and/or
the European Union.


\newpage



\newpage

\begin{figure}
\plotone{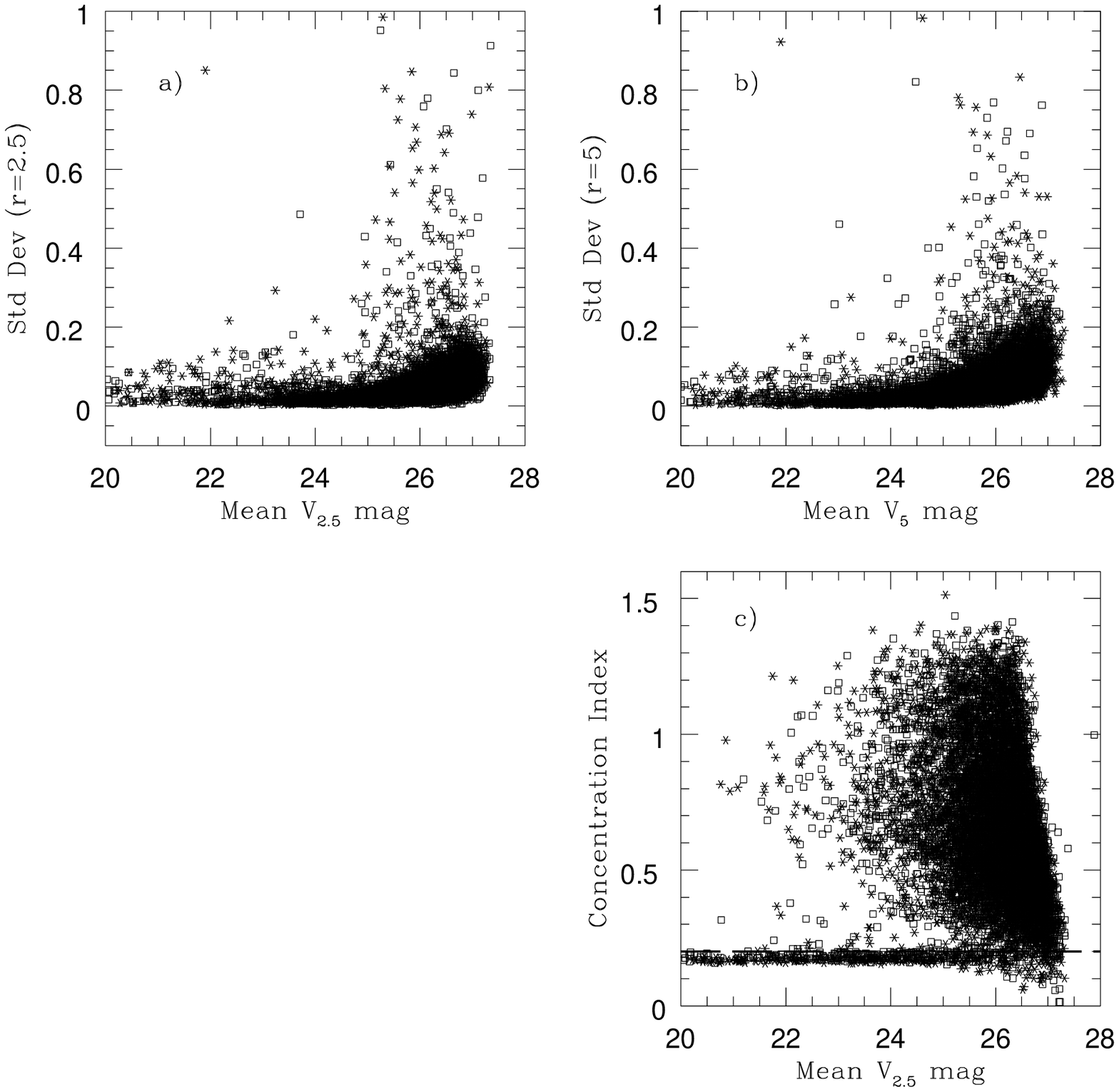}
\caption{Mean V magnitude vs. standard deviation for all sources in the the GOODS-N (squares)
and GOODS-S (asterisks) fields with a) r$=$2.5 pixel photometry and b) r$=$5 pixel photometry.
Average V magnitude vs. concentration index (V$_{2.5}$ - V$_{5}$) is shown in panel c, where
the dashed line represents the separation between resolved/extended sources and 
unresolved/point-like sources.  All sources have 5 magnitude measurements from which the
standard deviation was determined.}
\end{figure}

\begin{figure}
\plotone{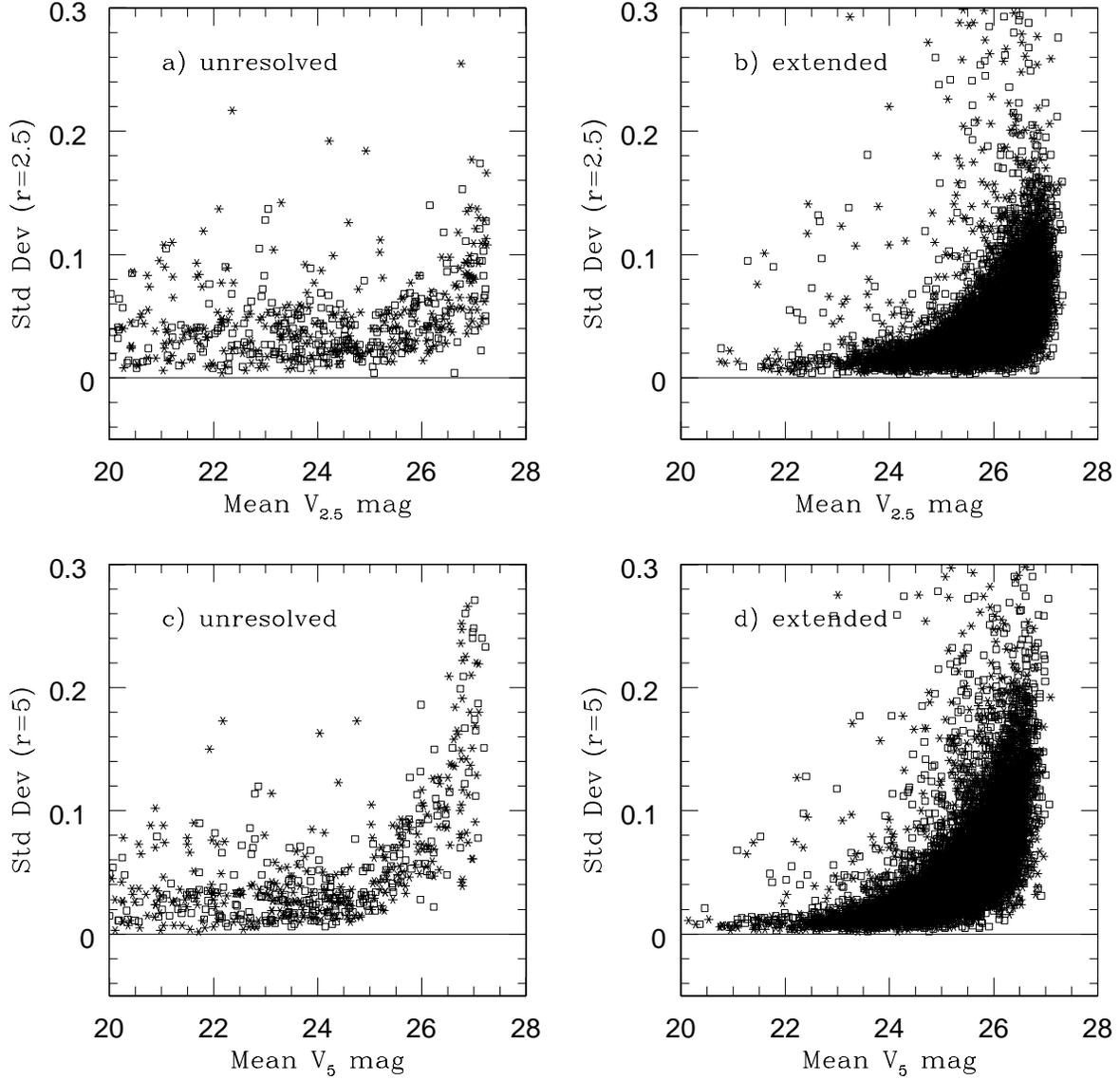}
\caption{Mean V magnitude vs. standard deviation for all sources in the the GOODS-N (squares)
and GOODS-S (asterisks) fields. a) unresolved sources and b) resolved sources shown using
r$=$2.5 pixel photometry.  c) unresolved sources and d) resolved sources shown using
r$=$5 pixel photometry.}
\end{figure}

\begin{figure}
\plotone{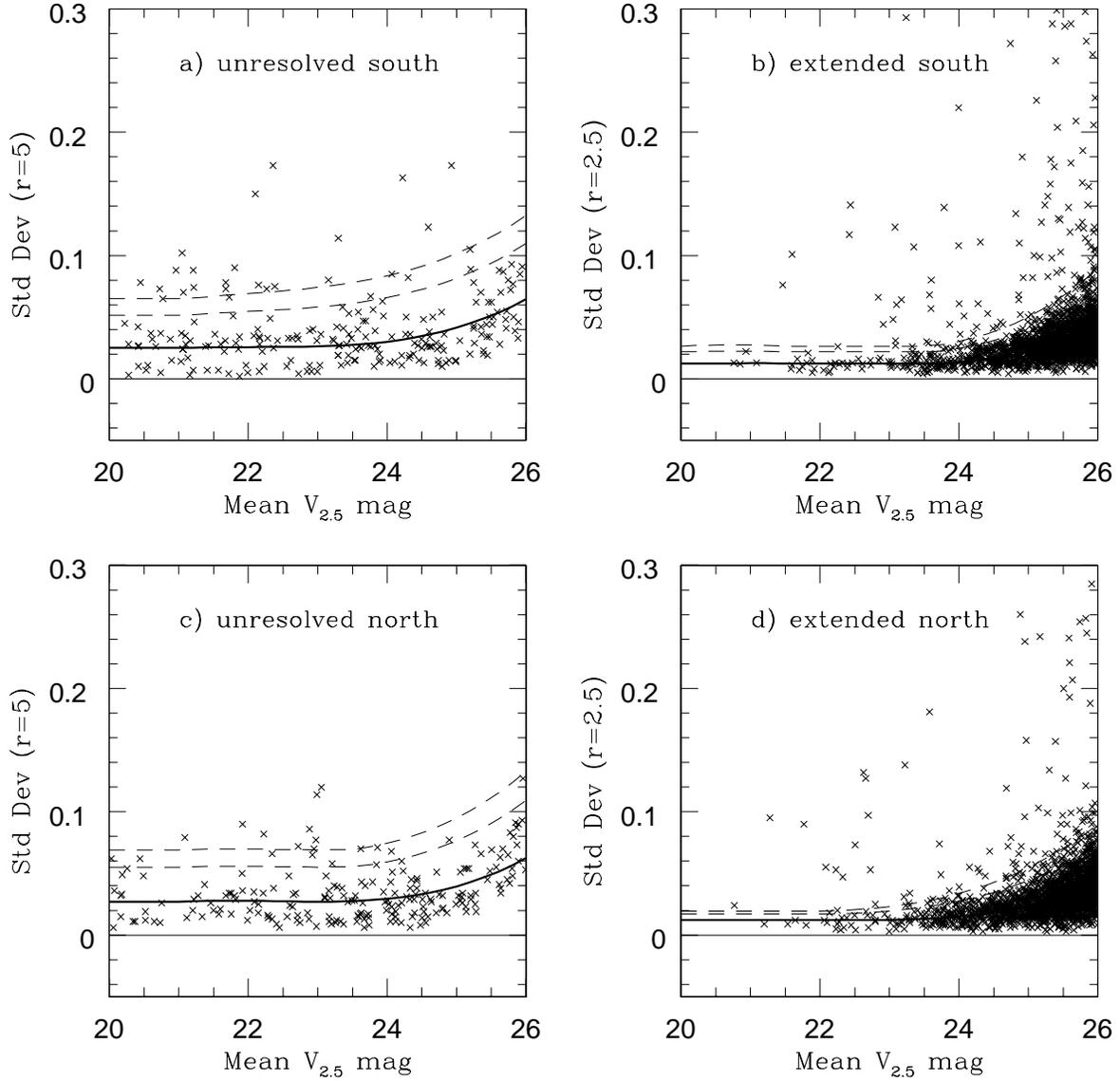}
\caption{Mean V magnitude vs. standard deviation for a) unresolved sources in GOODS-S,
b) extended sources in GOODS-S, c) unresolved sources in GOODS-N, d) extended sources
in GOODS-N.  Mean V magnitudes are measured within r$=$2.5 pixel apertures.
Standard deviation values are calculated using r$=$2.5 pixel apertures for extended sources and
r$=$5 pixel apertures for unresolved sources.
The solid line is a fit to the center of the distribution of standard deviation
values as a function of magnitude.
The dashed lines represent the mean plus 2$\sigma$ and mean plus 3$\sigma$
standard deviation values, where $\sigma$ is determined from the
Gaussian fit to the binned standard deviation distributions at each magnitude.}
\end{figure}

\begin{figure}
\plotone{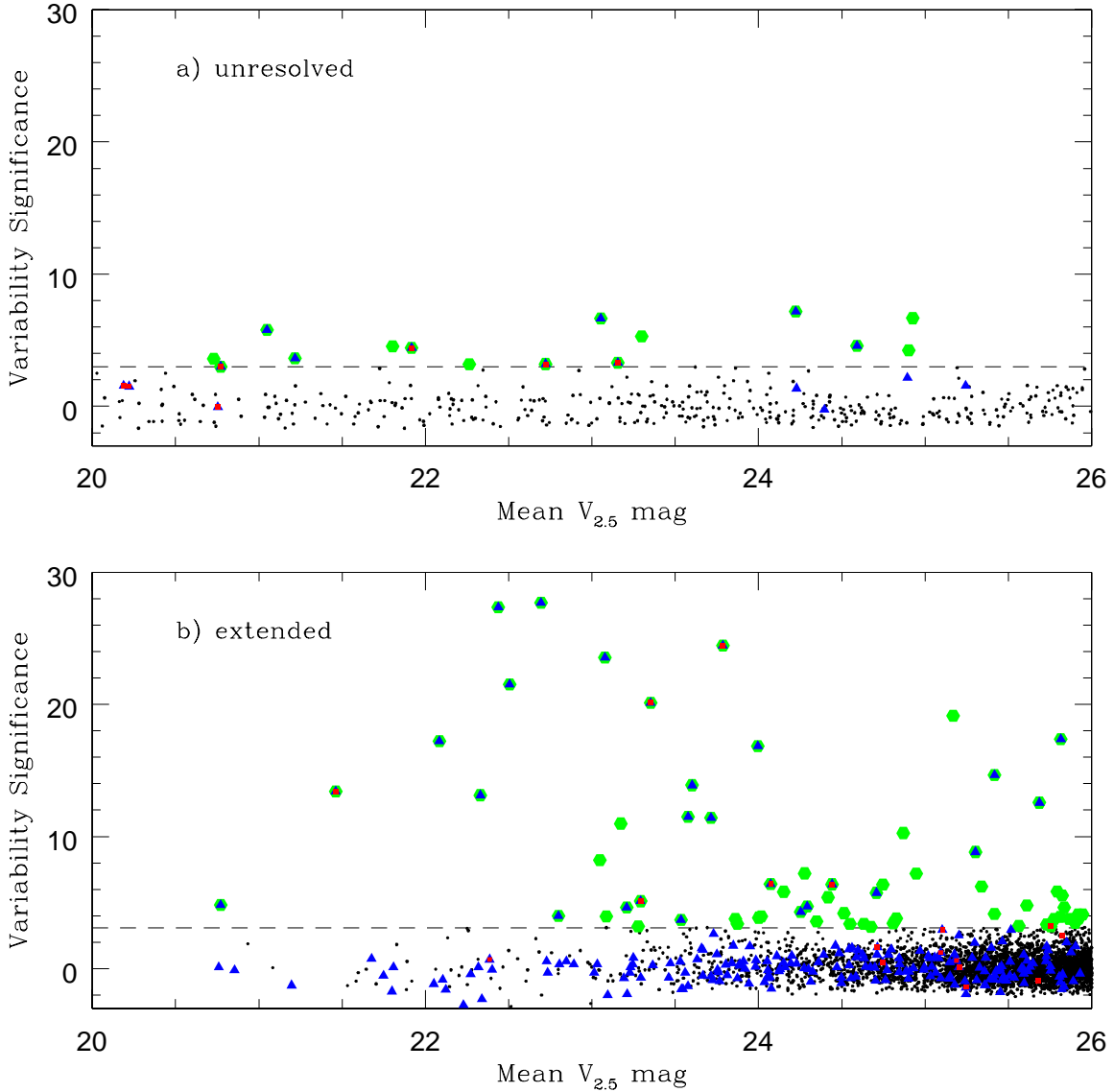}
\caption{Mean V magnitude vs. variability significance for
a) unresolved sources and b) extended sources.  The variability significance is the
standard deviation with the mean subtracted, divided by 1$\sigma$ from the Gaussian fit to binned
standard deviation distributions at each magnitude (see dashed lines in Figure 3).
The dashed lines represent the threshold for significant variability (see text and Figure 5).
Green points indicate all significantly varying sources
(above 3.1$\sigma$ for extended sources and 3.0$\sigma$ for unresolved sources).
Blue triangles indicate sources associated with X-ray emission in the 2Ms Chandra surveys.
Red squares represent Spitzer mid-IR power-law sources.}
\end{figure}

\begin{figure}
\plotone{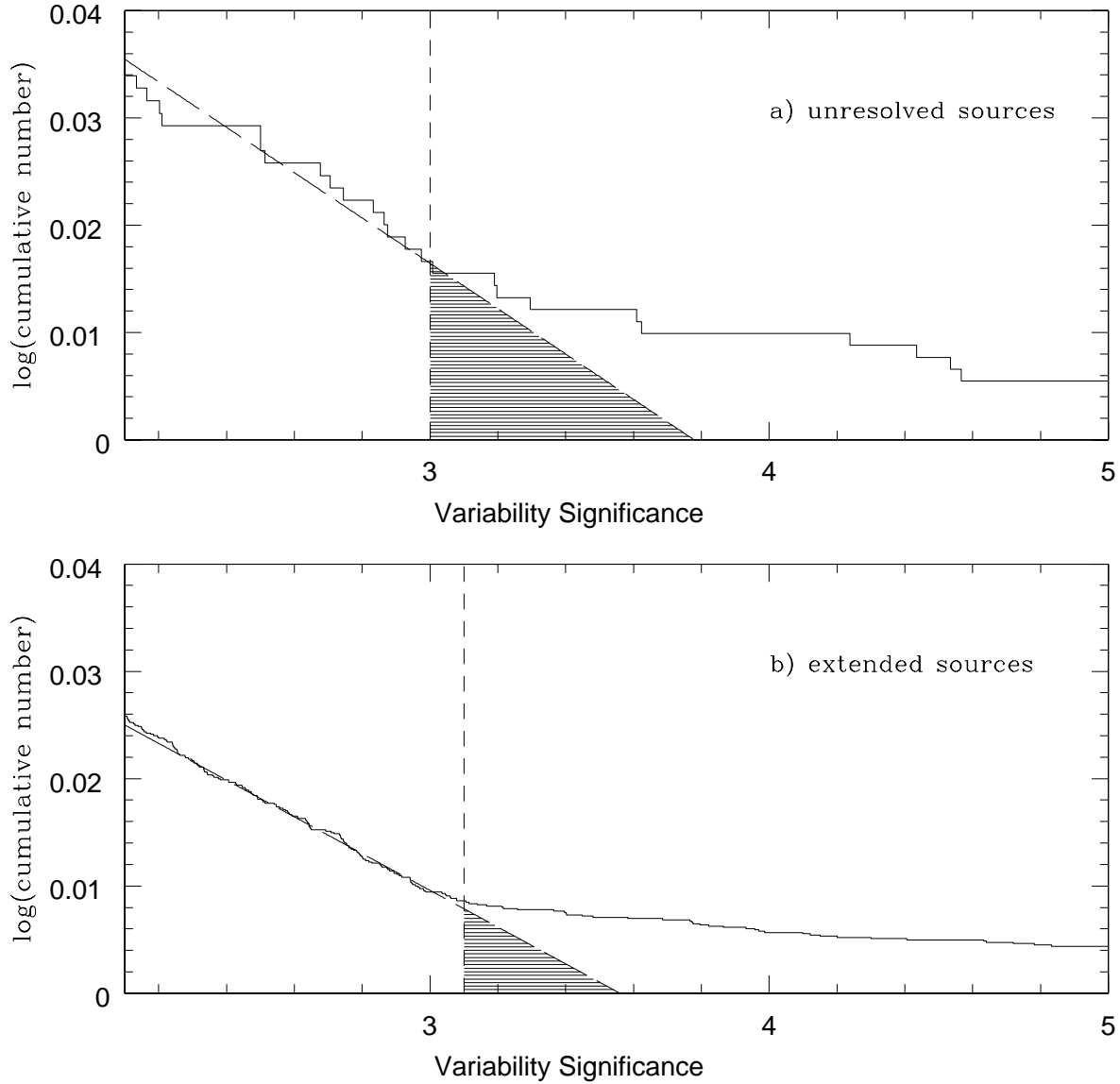}
\caption{Variability significance cumulative histogram for a) unresolved sources and b) extended sources
(solid line).  The long dashed line represents a fit to the distribution at low significance
values ($\lesssim$3).  The short dashed line is the approximate break of the distribution
from the low significance distribution fit (i.e. at values higher than this threshold, we
consider the source to be significantly varying beyond the normal galaxy population).
The shaded region represents the expected false positive population among the significantly
varying sample ($\sim$3 sources among the unresolved sources and $\sim$4 sources among the
extended sources).}
\end{figure}

\begin{figure}
\plotone{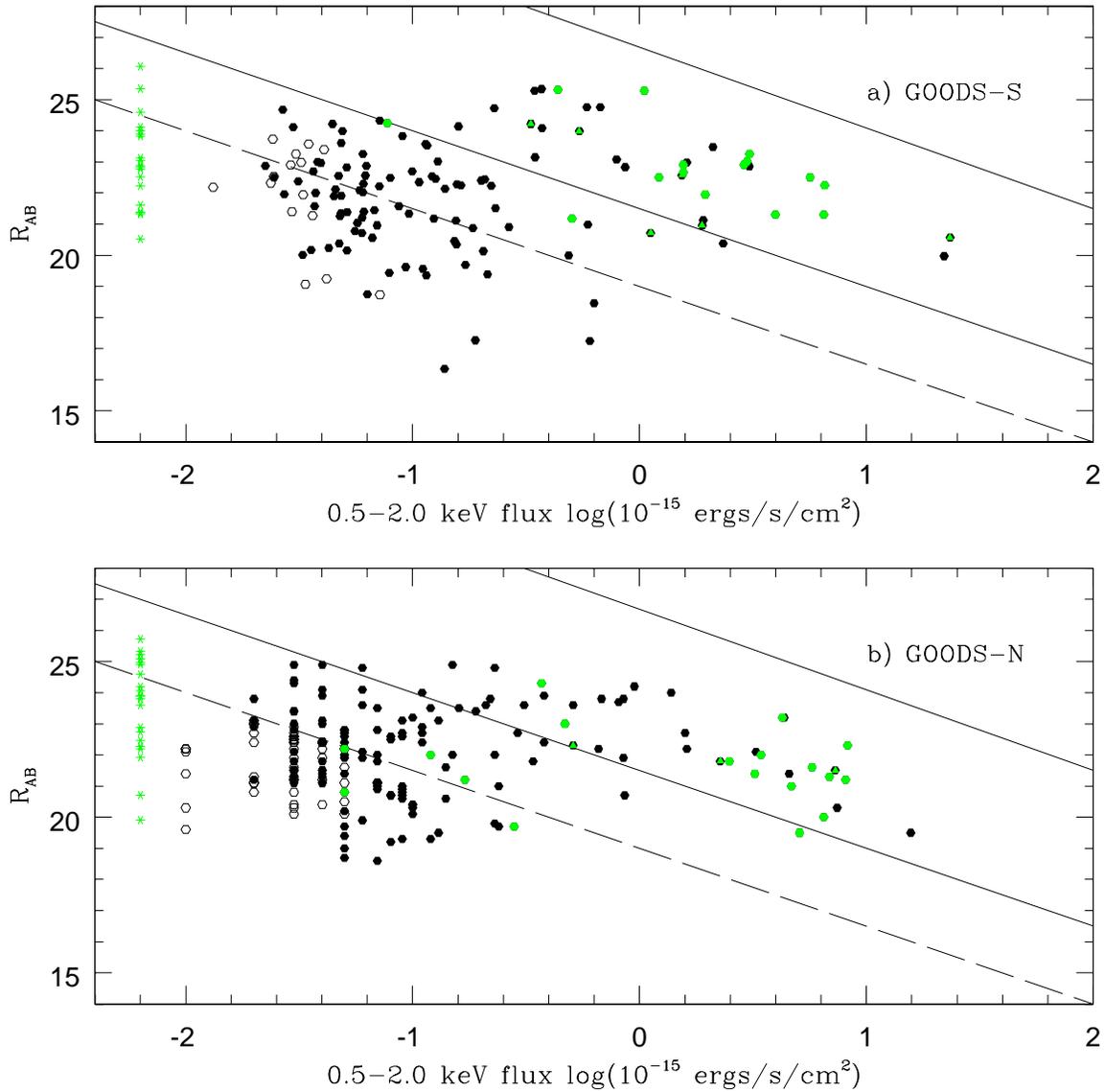}
\caption{Soft X-ray flux vs. R-band magnitude for objects in a) GOODS south and b) GOODS north.  Black points
are all X-ray sources with optical counterparts in the variability survey.  Green points are those
with significant variability (triangles are unresolved objects).
The green asterisks are upper limits for non-X-ray detected variables.
Solid lines represent log(F$_x$/F$_{opt}$) = -1, +1 and the dashed line is at -2.}
\end{figure}

\begin{figure}
\plotone{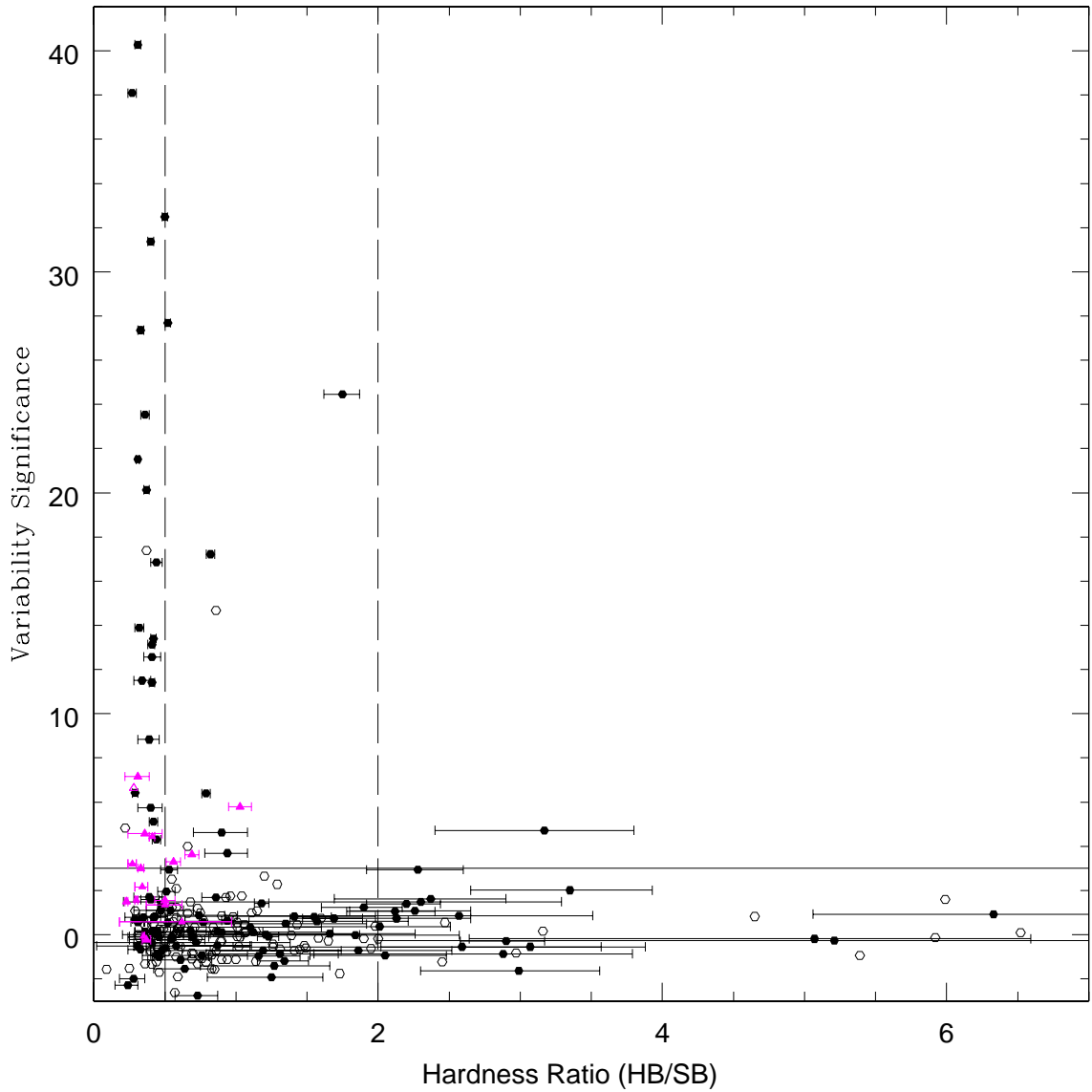}
\caption{Hardness ratio vs. variability significance for all X-ray sources with optical counterparts.
Hardness ratio is defined as the hard X-ray counts (2-7 keV) divided by the soft X-ray counts
(0.5-2 keV).  Open circles are objects with upper limits only.  Magenta triangles are unresolved
sources.  The dashed lines separate unobscured sources (HR$\leq$0.5) from moderately obscured sources
and obscured sources (HR$\geq$2).}
\end{figure}

\begin{figure}
\plotone{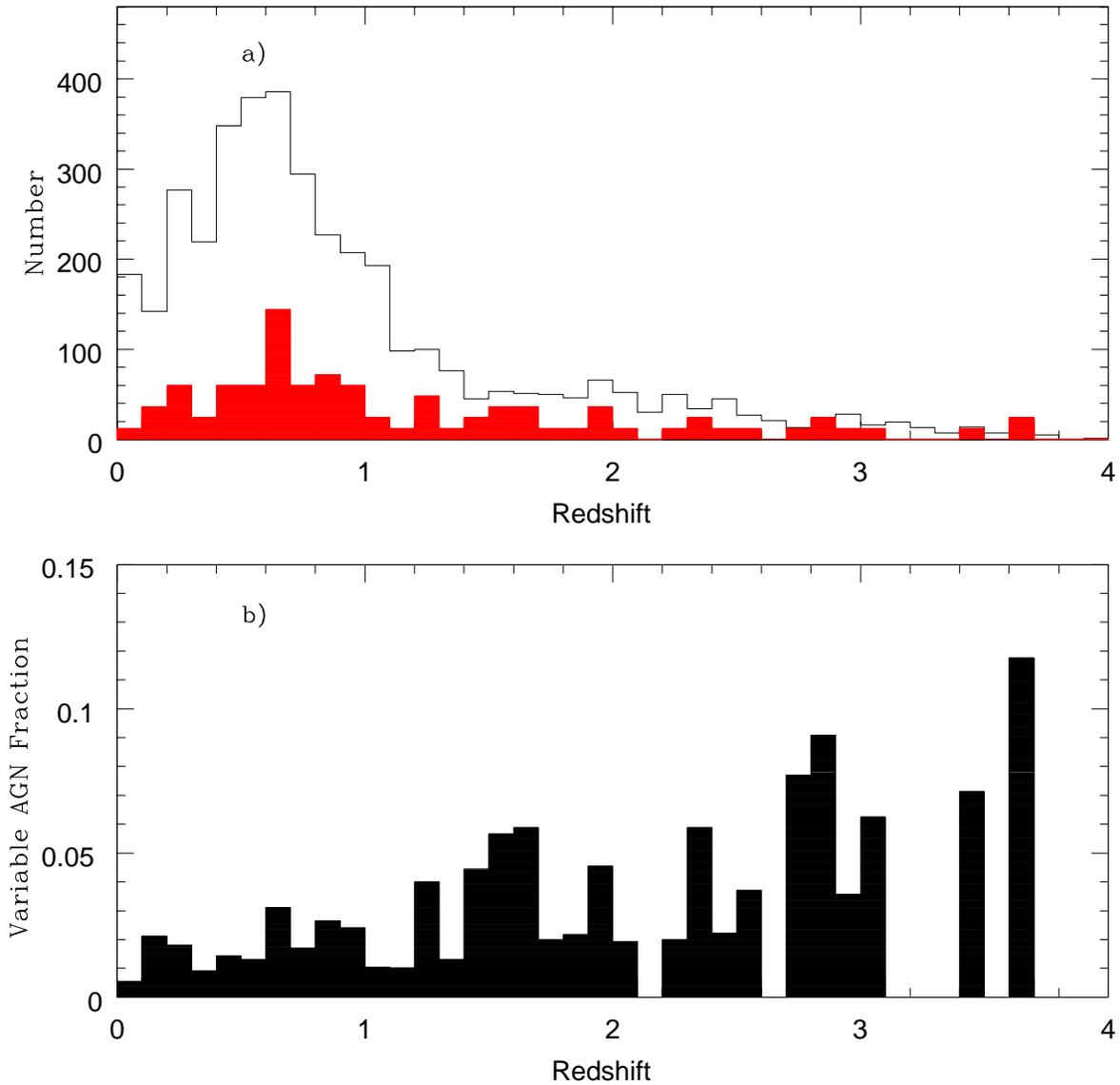}
\caption{a) Redshift distribution for galaxies in the GOODS variability survey (solid line) and for the significant
variables (red shaded histogram) plotted at a scale of 12:1. b) The fraction of galaxies containing
a varying nucleus as a function of redshift.}
\end{figure}

\begin{figure}
\plotone{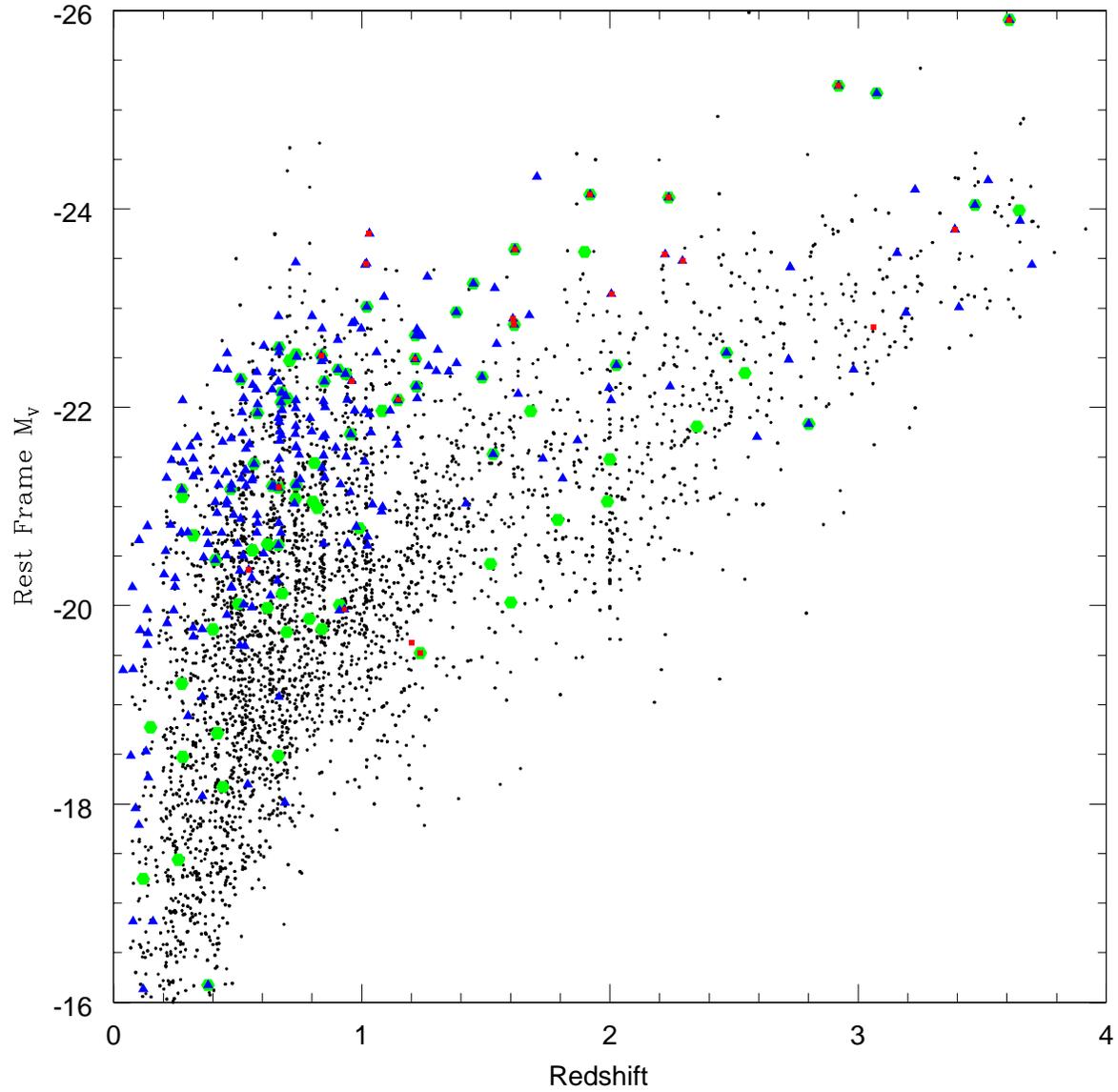}
\caption{Redshift vs. rest-frame absolute V magnitude for all galaxies (black points), optical 
variables (green circles),
X-ray sources (blue triangles) and mid-IR power-law sources (red squares).}
\end{figure}

\begin{figure}
\plotone{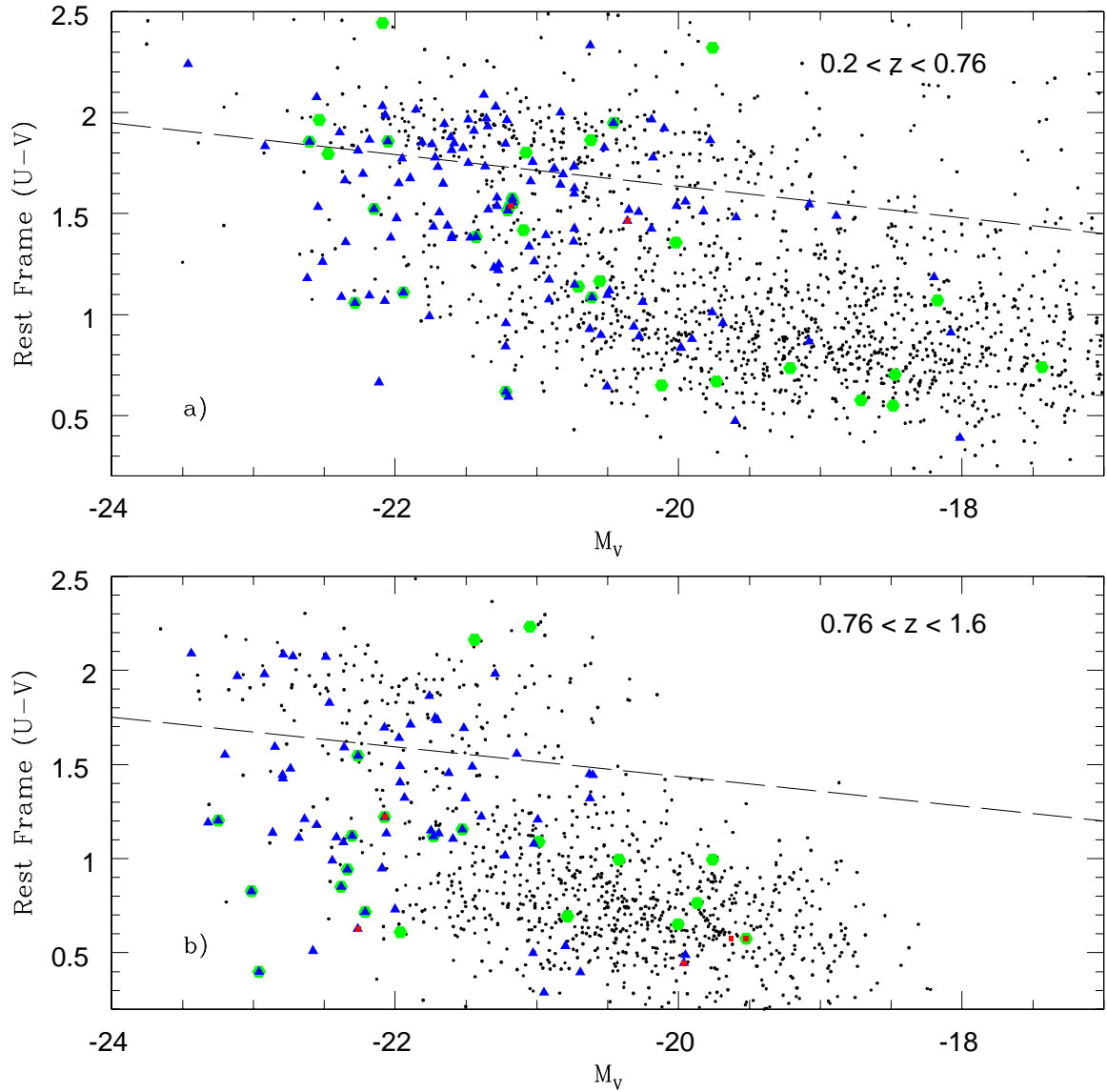}
\caption{Color-magnitude diagram for GOODS galaxies at a) 0.2$<$z$<$0.76 and b) 0.76$<$z$<$1.6.
Symbols are as in Fig 9.  Dashed lines represent the lower limit of red sequence galaxies
at the average redshift for each panel (Bell et al. 2004).}
\end{figure}

\begin{figure}
\plotone{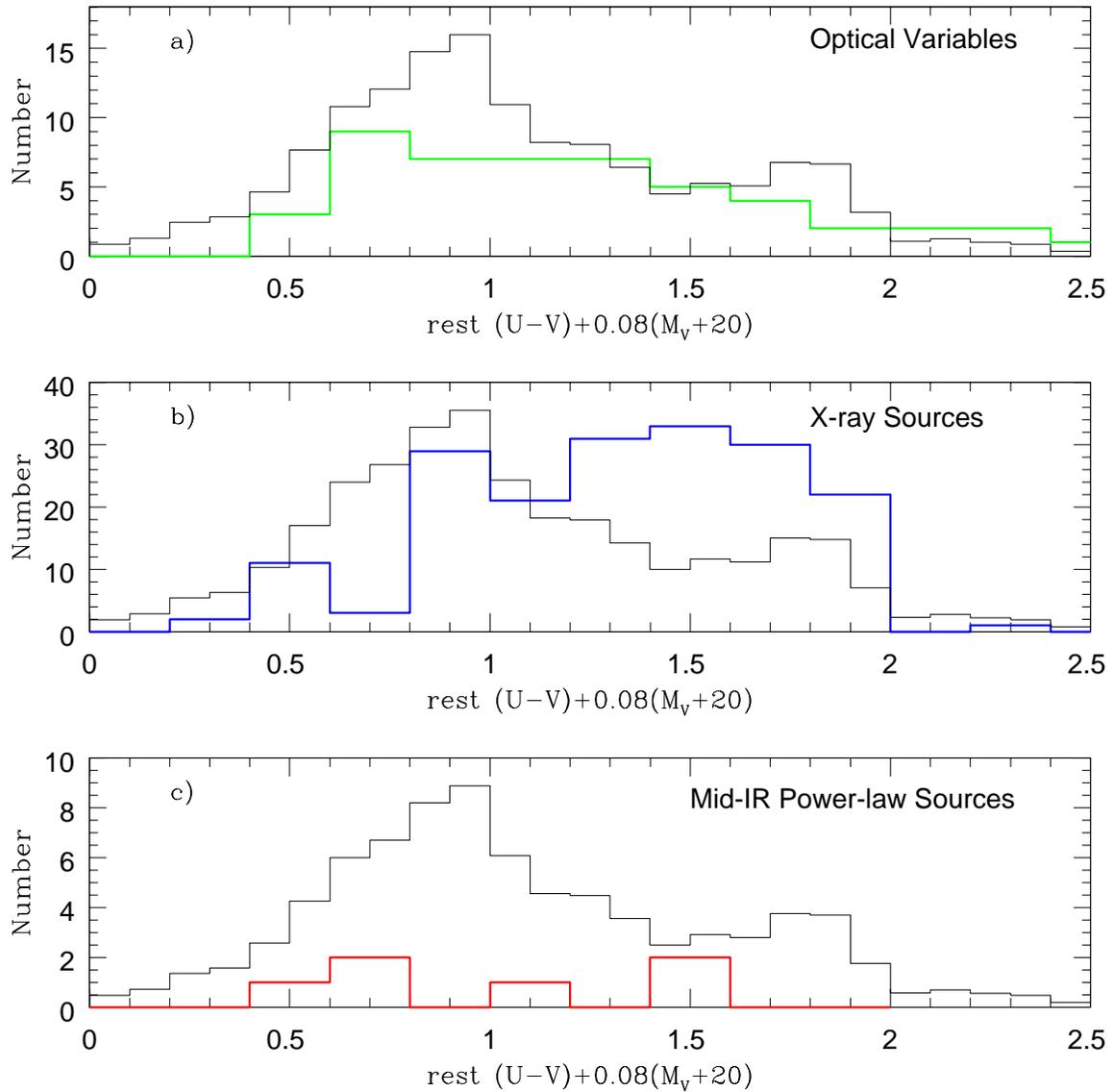}
\caption{Galaxy color histograms for all galaxies at z$<$1.6 in GOODS (solid line in all panels).
a) variability selected AGN hosts (green), b) X-ray selected AGN hosts (blue), and c) mid-IR power-law selected
AGN hosts at z$<$1.6.}
\end{figure}

\begin{figure}
\plotone{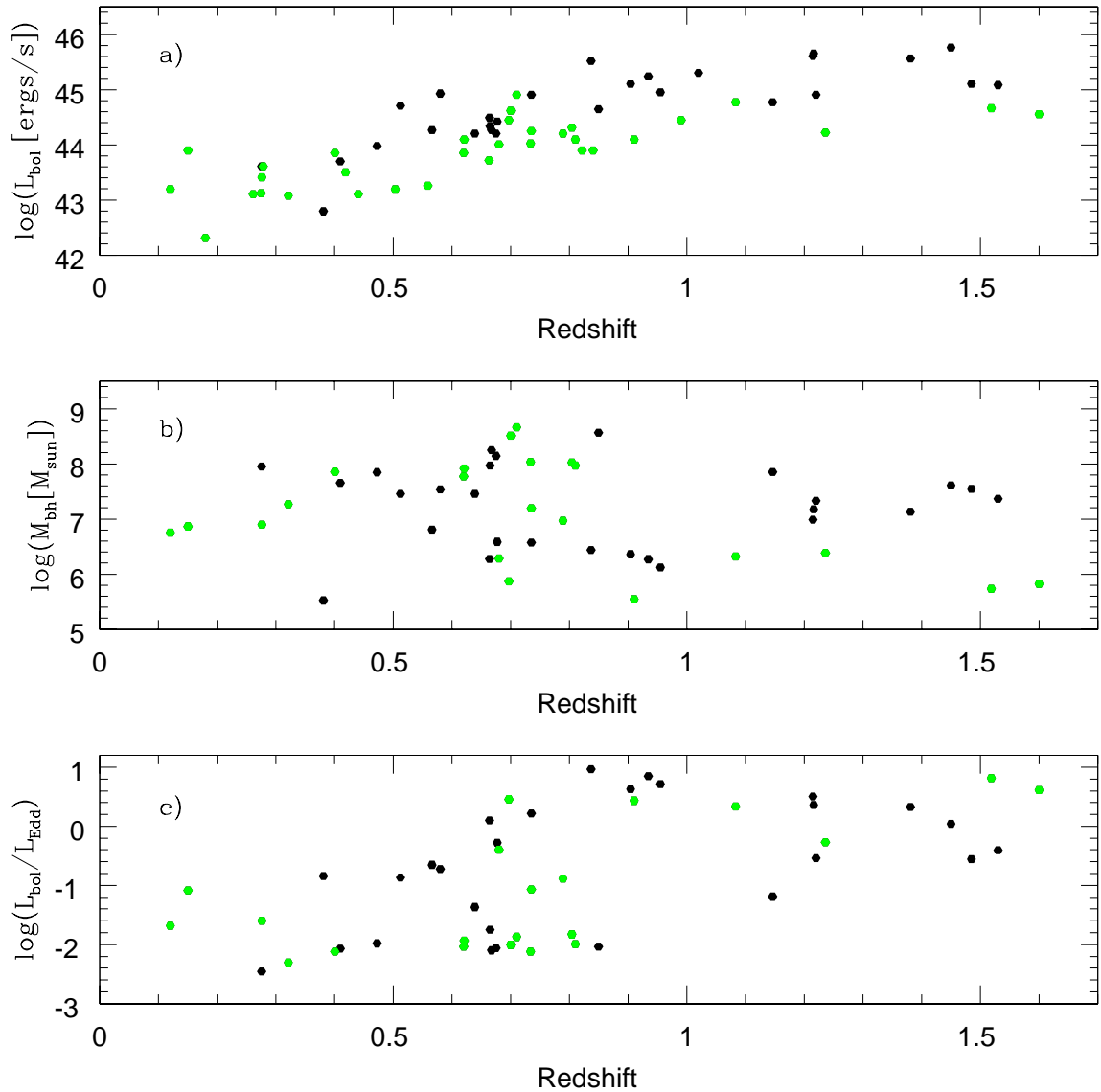}
\caption{a) Bolometric luminosity, b) black hole mass, and c) Eddington ratio vs redshift
for variable galaxies at z$<$1.6.  Black points are variables that are also X-ray
sources and green points indicate non-X-ray detected variables.}
\end{figure}

\begin{deluxetable}{cllcccccccc}
\tabletypesize{\scriptsize}
\tablecaption{Variable Nuclei Galaxies in GOODS}
\tablewidth{0pt}
\setlength{\tabcolsep}{0.04in}
\tablehead{
\colhead{GOODS ID} & \colhead{RA} & \colhead{DEC} & \colhead{V$_{2.5}$} & \colhead{CI} &
\colhead{Std Dev} & \colhead{Significance} & \colhead{Redshift} & \colhead{ph/sp} &
\colhead{Note} & \colhead{X-ray/IR AGN}}
\startdata
  263 & 188.9272156 & 62.2081223 & 25.730 & 0.946 & 0.084  & 3.365 & 0.440 & ph & 6 & -- \\
 1177 & 188.9713287 & 62.1769981 & 22.330 & 0.210 & 0.047  & 3.127 & 1.381 & sp & 2 & X-ray \\
 3164 & 189.0242615 & 62.1966934 & 24.708 & 0.418 & 0.072  & 5.745 & 1.485 & sp & 2 & X-ray \\
 6370 & 189.0749664 & 62.2763901 & 21.770 & 0.242 & 0.090  & 2.488 & 0.580 & ph & 6 & X-ray \\
 6527 & 189.0774994 & 62.1875343 & 22.506 & 0.214 & 0.073  & 1.511 & 1.020 & sp & 1 & X-ray \\
 7321 & 189.0874634 & 62.2367172 & 25.302 & 0.298 & 0.134  & 8.833 & 1.530 & ph & 6 & X-ray \\
 8848 & 189.1068878 & 62.2799873 & 24.826 & 0.350 & 0.058  & 3.821 & 2.000 & ph & 6 & -- \\
 9104 & 189.1100006 & 62.1646309 & 25.776 & 0.924 & 0.092  & 3.773 & 0.840 & sp & 1 & -- \\
 9345 & 189.1127319 & 62.2994232 & 24.676 & 0.660 & 0.048  & 3.166 & 1.680 & ph & 6 & -- \\
 9386 & 189.1133118 & 62.1260033 & 24.746 & 0.972 & 0.079  & 6.356 & 0.321 & sp & 1 & -- \\
10066 & 189.1220245 & 62.2704849 & 24.292 & 0.676 & 0.051  & 4.720 & 0.850 & sp & 1 & X-ray \\
11426 & 189.1384735 & 62.1429482 & 22.658 & 0.314 & 0.127  & 38.092 & 0.934 & sp & 2 & X-ray \\
12198 & 189.1471252 & 62.1860580 & 23.208 & 0.638 & 0.031  & 4.633 & 0.409 & sp & 1 & X-ray \\
13302 & 189.1590424 & 62.1154747 & 25.564 & 0.672 & 0.076  & 3.258 & 3.650 & ph & 6 & -- \\
14163 & 189.1684875 & 62.1062279 & 25.864 & 0.660 & 0.096  & 3.776 & 2.350 & ph & 6 & -- \\
15439 & 189.1825867 & 62.1221237 & 25.918 & 0.748 & 0.104  & 4.121 & 1.990 & ph & 6 & -- \\
16346 & 189.1929932 & 62.2707367 & 25.812 & 1.274 & 0.096  & 3.931 & 0.503 & sp & 2 & -- \\
18014 & 189.2113037 & 62.1388969 & 25.824 & 0.458 & 0.121  & 5.531 & 1.519 & sp & 1 & -- \\
21365 & 189.2451630 & 62.2430305 & 23.716 & 0.528 & 0.074  & 11.415 & 0.677 & sp & 2 & X-ray \\
24210 & 189.2760620 & 62.3601723 & 22.628 & 0.228 & 0.132  & 40.276 & 0.904 & sp & 1 & X-ray \\
24637 & 189.2812042 & 62.3632812 & 23.220 & 0.232 & 0.138  & 31.371 & 1.450 & sp & 2 & X-ray \\
25631 & 189.2930145 & 62.3034363 & 24.002 & 0.350 & 0.039  & 3.865 & 0.697 & sp & 1 & -- \\
26715 & 189.3054657 & 62.3478355 & 25.612 & 0.718 & 0.099  & 4.781 & 0.180 & ph & 6 & -- \\
26774 & 189.3061218 & 62.3603401 & 24.350 & 1.102 & 0.044  & 3.583 & 0.275 & sp &  -& - \\
27941 & 189.3194733 & 62.2925835 & 24.442 & 0.360 & 0.068  & 6.385 & 1.146 & sp & 1 & X-ray,IR \\
28356 & 189.3245087 & 62.3154716 & 23.292 & 0.254 & 0.034  & 5.112 & 2.237 & sp & 2 & X-ray,IR \\
30926 & 189.3559875 & 62.2489471 & 25.922 & 0.346 & 0.099  & 3.795 & 2.545 & sp & 3 & -- \\
30987 & 189.3569031 & 62.2801285 & 20.770 & 0.316 & 0.024  & 4.832 & 0.000 & sp & 1 & X-ray \\
32037 & 189.3706360 & 62.1909637 & 25.168 & 0.544 & 0.242  & 19.125 & 0.810 & ph & 6 & -- \\
32542 & 189.3768311 & 62.2312050 & 25.416 & 0.334 & 0.082  & 4.150 & 1.600 & ph & 6 & -- \\
34017 & 189.3970337 & 62.3355713 & 24.152 & 0.644 & 0.055  & 5.822 & 1.083 & sp & 2 & -- \\
35004 & 189.4146118 & 62.2506256 & 25.836 & 1.108 & 0.108  & 4.640 & 0.821 & sp & 1 & -- \\
 2426 & 189.0090942 & 62.2106133 & 23.862 & 0.560 & 0.036  & 3.767 & 0.639 & sp & 2 & X-ray \\
 4680 & 189.0501404 & 62.1941376 & 22.694 & 0.934 & 0.097  & 27.694 & 0.276 & sp & 2 & X-ray \\
 4809 & 189.0520020 & 62.1945724 & 23.279 & 0.724 & 0.026  & 3.216 & 0.276 & sp & 1 & -- \\
 7902 & 189.0950012 & 62.2166138 & 22.798 & 0.652 & 0.025  & 3.987 & 0.473 & sp & 2 & X-ray \\
10848 & 189.1320648 & 62.1311073 & 24.948 & 0.426 & 0.096  & 7.200 & 1.790 & ph & 6 & -- \\
17767 & 189.2086639 & 62.1337738 & 25.900 & 1.248 & 0.093  & 3.480 & 0.559 & sp & 1 & -- \\
22976 & 189.2613525 & 62.2621117 & 22.084 & 0.378 & 0.055  & 17.224 & 0.512 & sp & 2 & X-ray \\
23832 & 189.2719421 & 62.2962265 & 24.810 & 0.558 & 0.054  & 3.448 & 0.789 & sp & 2 & -- \\
 1667 & 188.9900970 & 62.1734390 & 23.052 & 0.196 & 0.120  & 6.635 & 3.075 & sp & 2 & X-ray \\
35757 & 189.4272003 & 62.3033028 & 21.918 & 0.180 & 0.090  & 4.435 & 2.309 & sp & 2 & X-ray,IR \\
38313 & 189.4888000 & 62.2742996 & 22.722 & 0.178 & 0.072  & 3.197 & 2.922 & sp & 1 & X-ray,IR \\
  820 & 53.0084343 -& 27.7590332 & 25.946 & 0.616 & 0.105  & 4.099 & 1.899 & ph & 4 & -- \\
 2967 & 53.0344429 -& 27.6982098 & 25.816 & 0.246 & 0.298  & 17.375 & 2.470 & sp & 4 & X-ray \\
 4234 & 53.0454674 -& 27.7374840 & 23.352 & 0.256 & 0.107  & 20.128 & 1.615 & sp & 4 & X-ray,IR \\
 6038 & 53.0586662 -& 27.7084370 & 25.418 & 0.226 & 0.204  & 14.668 & 2.026 & sp & 5 & X-ray \\
 6254 & 53.0601578 -& 27.7734680 & 24.416 & 0.544 & 0.053  & 5.406 & 0.734 & sp & 4 & -- \\
 6595 & 53.0624199 -& 27.8575153 & 24.254 & 0.622 & 0.042  & 4.300 & 0.675 & sp & 4 & X-ray \\
 8830 & 53.0767479 -& 27.8174648 & 24.548 & 0.818 & 0.042  & 3.396 & 0.261 & sp & 5 & -- \\
 9055 & 53.0782013 -& 27.8784523 & 25.794 & 0.790 & 0.122  & 5.831 & 0.910 & ph & 4 & -- \\
10019 & 53.0842857 -& 27.8140373 & 24.018 & 0.808 & 0.036  & 3.969 & 0.735 & sp & 4 & -- \\
11674 & 53.0942993 -& 27.8534985 & 24.868 & 0.774 & 0.110  & 10.266 & 0.990 & ph & 4 & -- \\
13494 & 53.1048546 -& 27.7052193 & 24.072 & 0.252 & 0.051  & 6.415 & 1.617 & sp & 4 & X-ray,IR \\
14049 & 53.1079750 -& 27.7337475 & 23.086 & 0.530 & 0.031  & 3.957 & 0.278 & sp & 4 & -- \\
15416 & 53.1146431 -& 27.9367771 & 25.338 & 0.612 & 0.098  & 6.226 & 0.663 & sp & 5 & -- \\
17366 & 53.1249123 -& 27.7583027 & 21.462 & 0.202 & 0.076  & 13.403 & 1.215 & sp & 4 & X-ray,IR \\
17438 & 53.1252556 -& 27.7565346 & 23.602 & 0.230 & 0.080  & 13.885 & 0.955 & sp & 4 & X-ray \\
17558 & 53.1259003 -& 27.7512760 & 22.438 & 0.214 & 0.141  & 27.362 & 0.735 & sp & 4 & X-ray \\
18648 & 53.1315727 -& 27.7702045 & 24.634 & 0.840 & 0.044  & 3.402 & 0.680 & ph & 4 & -- \\
18812 & 53.1324959 -& 27.8528843 & 23.048 & 0.202 & 0.051  & 8.213 & 0.400 & ph & 4 & -- \\
19698 & 53.1375771 -& 27.7001095 & 25.754 & 0.318 & 0.081  & 3.211 & 1.236 & sp & 5 & IR \\
23092 & 53.1560745 -& 27.6666927 & 23.578 & 0.288 & 0.068  & 11.501 & 0.664 & sp & 4 & X-ray \\
24366 & 53.1636200 -& 27.7053509 & 24.276 & 0.410 & 0.061  & 7.212 & 1.764 & sp & 5 & -- \\
26093 & 53.1744499 -& 27.7332993 & 25.686 & 0.222 & 0.209  & 12.566 & 0.381 & sp & 5 & X-ray \\
26882 & 53.1801491 -& 27.8206043 & 23.240 & 0.230 & 0.293  & 59.702 & 1.920 & sp & 4 & X-ray,IR \\
29404 & 53.2007370 -& 27.8823910 & 23.534 & 0.614 & 0.030  & 3.686 & 0.667 & sp & 4 & X-ray \\
29808 & 53.2045441 -& 27.8972778 & 24.512 & 0.702 & 0.047  & 4.200 & 0.419 & sp & 4 & -- \\
31429 & 53.2203522 -& 27.8555088 & 23.996 & 0.332 & 0.108  & 16.851 & 1.220 & sp & 4 & X-ray \\
 7929 & 53.0714340 -& 27.7175846 & 23.078 & 0.736 & 0.123  & 23.532 & 0.566 & sp & 4 & X-ray \\
15508 & 53.1150970 -& 27.6958046 & 23.786 & 0.516 & 0.139  & 24.456 & 0.665 & sp & 4 & X-ray,IR \\
15694 & 53.1161232 -& 27.8933048 & 23.872 & 0.270 & 0.031  & 3.401 & 0.621 & sp & 5 & -- \\
18828 & 53.1325912 -& 27.8529434 & 23.172 & 0.212 & 0.064  & 10.979 & 2.796 & sp & 5 & -- \\
 1872 & 53.0232468 -& 27.7772713 & 22.264 & 0.196 & 0.073  & 3.190 & 0.710 & ph & 4 & -- \\
 8701 & 53.0760040 -& 27.8781586 & 24.592 & 0.194 & 0.123  & 4.567 & 2.801 & sp & 4 & X-ray \\
14887 & 53.1119652 -& 27.7132568 & 24.902 & 0.198 & 0.124  & 4.238 & 0.804 & ph & 4 & -- \\
22530 & 53.1528091 -& 27.8778725 & 24.926 & 0.174 & 0.173  & 6.670 & 0.620 & ph & 4 & -- \\
23424 & 53.1580276 -& 27.7691936 & 20.728 & 0.162 & 0.073  & 3.609 & 0.150 & ph & 4 & -- \\
23556 & 53.1588287 -& 27.6624470 & 20.772 & 0.164 & 0.065  & 3.008 & 0.837 & sp & 4 & X-ray,IR \\
24251 & 53.1628609 -& 27.7671623 & 21.218 & 0.158 & 0.074  & 3.624 & 1.216 & sp & 4 & X-ray \\
26077 & 53.1743851 -& 27.8673534 & 23.156 & 0.178 & 0.080  & 3.296 & 3.610 & sp & 4 & X-ray,IR \\
26512 & 53.1775360 -& 27.9149456 & 21.804 & 0.186 & 0.090  & 4.534 & 0.120 & ph & 4 & -- \\
27457 & 53.1846352 -& 27.8809185 & 24.224 & 0.188 & 0.163  & 7.162 & 3.471 & sp & 4 & X-ray \\
29365 & 53.2003708 -& 27.7898388 & 23.298 & 0.182 & 0.114  & 5.287 & 0.700 & ph & 4 & -- \\
 3485 & 53.0393639 -& 27.8018875 & 21.048 & 0.168 & 0.102  & 5.786 & 2.810 & sp & 4 & X-ray \\
\enddata
\tablecomments{1 - Barger et al. 2008, 2 - Wirth et al. 2008, 3 - Reddy et al. 2006, 4 - Barro et al.
2010, 5 - Popesso et al. 2009, 6 - Bundy et al. 2009}

\end{deluxetable}

\begin{deluxetable}{ccccc}
\tabletypesize{\scriptsize}
\tablecaption{X-Ray Stacking Results}
\tablewidth{0pt}
\setlength{\tabcolsep}{0.07in}
\tablehead{
\colhead{Energy Band (keV)} & \colhead{No. Stacked Sources} & \colhead{Total Counts}\tablenotemark{1} & 
\colhead{Total Background Counts}\tablenotemark{2} &
\colhead{S/N}\tablenotemark{3}}

\startdata
\cutinhead{GOODS-N}
0.5 - 0.7 & 20 & 228 $\pm$ 15 & 205.1 $\pm$ 0.5 & 1.5 \\
0.5 - 2.0 & 20 &  96 $\pm$ 10 &  67.7 $\pm$ 0.3 & 2.9 \\
2.0 - 7.0 & 19 & 127 $\pm$ 11 & 130.3 $\pm$ 0.4 & -0.3 \\
\cutinhead{GOODS-S}
0.5 - 0.7 & 19 & 273 $\pm$ 17 & 234.1 $\pm$ 0.5 & 2.4 \\
0.5 - 2.0 & 20 &  99 $\pm$ 10 &  78.4 $\pm$ 0.3 & 2.1 \\
2.0 - 7.0 & 18 & 170 $\pm$ 13 & 152.0 $\pm$ 0.4 & 1.4 \\
\enddata
\tablenotetext{1}{Total counts extracted within a radius r=1.75$\arcsec$ of source location.}
\tablenotetext{2}{Based on 1000 samples taken within an annulus r=10$\arcsec$ to r=30$\arcsec$
around source position.}
\tablenotetext{3}{S/N = S/(sqrt(S+B)), where S is net source count flux per source and B is
background count flux per source.}

\end{deluxetable}

\end{document}